\documentclass[11pt]{article}
\usepackage{rotating}
\usepackage{epsf, epsfig}

\def\prd{{Phys.~Rev.~D}}

\def\apj{{Ap.~J.}}

\def\cc{{\rm\, cm}^{-3}}
\def\hmpc{{\rm\, h^{-1}Mpc}}

\def\kms{{\rm\, km\ s^{-1}}}

\def\mpc{{\rm\,Mpc}}

\def\bx{{\bf x}}

\def\eg{{\it e.g.,\,}}
\def\ie{{\it i.e.,\,}}
\def\etal{{\it et al.\,}}
\def\et{{\it et al.\,}} 
\def\etc{{\it etc.\,}}

\def\p3m{P$^3$M}

\def\mpl{m_{\cal P}}

\def\spose#1{\hbox to 0pt{#1\hss}}
\def\lta{\mathrel{\spose{\lower 3pt\hbox{$\mathchar"218$}}
     \raise 2.0pt\hbox{$\mathchar"13C$}}}
\def\gta{\mathrel{\spose{\lower 3pt\hbox{$\mathchar"218$}}
     \raise 2.0pt\hbox{$\mathchar"13E$}}}
\def\simgt{\gta}
\def\simlt{\lta}

\def\eqright{\begin{eqnarray}}
\def\endeqright{\end{eqnarray}}

\def\be{\begin{equation}}
\def\ee{\end{equation}}
\def\bea{\begin{eqnarray}}

\def\eea{\end{eqnarray}}

\def\citet{\cite}
\def\citep{\cite}

\def\eg{{e.g., }}
\def\ie{{i.e., }}
\def\etal{{et al., }}
\def\et{{et al. }}
\def\etc{{etc.}}

\def\'{^{\prime}}

\def\mpl{m_{\cal P}}

\def\cc{{\rm~cm}^{-3}}
\def\hmpc{{\, {\rm h}^{-1}~\rm Mpc}}

\def\kms{{\rm~km~s^{-1}}}

\def\mpc{{\rm~Mpc}}

\def\spose#1{\hbox to 0pt{#1\hss}}
\def\lta{\mathrel{\spose{\lower 3pt\hbox{$\mathchar"218$}}
     \raise 2.0pt\hbox{$\mathchar"13C$}}}
\def\gta{\mathrel{\spose{\lower 3pt\hbox{$\mathchar"218$}}
     \raise 2.0pt\hbox{$\mathchar"13E$}}}
\def\ge{\mathrel{\spose{\lower 3pt\hbox{$-$}}
     \raise 2.0pt\hbox{$\mathchar"13E$}}}
\def\le{\mathrel{\spose{\lower 3pt\hbox{$-$}}
     \raise 2.0pt\hbox{$\mathchar"13C$}}}
\def\simgt{\gta}
\def\simlt{\lta}

\def\be{\begin{equation}}
\def\ee{\end{equation}}
\def\bea{\begin{eqnarray}}
\def\eea{\end{eqnarray}}

\def\etal{{\it et al. }\rm}
\def\simlt{\mathrel{\hbox{\rlap{\hbox{\lower4pt\hbox{$\sim$}}}\hbox{$<$}}}}
\def\simgt{\mathrel{\hbox{\rlap{\hbox{\lower4pt\hbox{$\sim$}}}\hbox{$>$}}}}
\newcounter{parentequation}

\def\beq{\begin{equation}}
\def\eeq{\end{equation}}

\def\ga{\mathrel{\raise.3ex\hbox{$>$\kern-.75em\lower1ex\hbox{$\sim$}}}}
\def\la{\mathrel{\raise.3ex\hbox{$<$\kern-.75em\lower1ex\hbox{$\sim$}}}}
\def\gev{{\rm \, Ge\kern-0.125em V}}
\def\tev{{\rm \, Te\kern-0.125em V}}
\def\gyr{{\rm \, G\kern-0.125em yr}}

\def\gappeq{\mathrel{\rlap {\raise.5ex\hbox{$>$}}
{\lower.5ex\hbox{$\sim$}}}}
\def\lappeq{\mathrel{\rlap{\raise.5ex\hbox{$<$}}
{\lower.5ex\hbox{$\sim$}}}}
\def\Toprel#1\over#2{\mathrel{\mathop{#2}\limits^{#1}}}

\def\mpl{M_{\rm Pl}}

\def\Kuo02{{Kuo \et\ 2003}}
\def\Goldstein02{{Goldstein \et\ 2003}}
\def\capp00{{Bond \et\ 2000}}
\def\bmrst02{{Bond \et\ 2002}}

\def\bjkrad00{{Bond, Jaffe \& Knox 2000}}
\def\bjkquad98{{Bond, Jaffe \& Knox 1998}}
\def\bj98{{Bond \& Jaffe 1999}}
\def\toco99{{Miller \et\ 1999}}
\def\mauskopf99{{Mauskopf \et\ 2000}}
\def\deBnature00{{de Bernardis \et\ 2000}}
\def\lange00{{Lange \et\ 2001}}  
\def\jaffe00{{Jaffe \et\ 2001}}  
\def\maxhanany00{{Hanany \et\ 2000}}
\def\nett01{{Netterfield \et\ 2002}}
\def\deBpkdip01{{de Bernardis \et\ 2002}}
\def\maxlee01{{Lee \et\ 2002}}
\def\Ruhl02{{Ruhl \et\ 2003}}
\def\DASI01{{Halverson \et\ 2002}}
\def\Mason02b{{Mason \et\ 2003}}
\def\Pearson02{{Pearson \et\ 2003}}
\def\Myers02{{Myers \et\ 2003}}
\def\Sievers02{{Sievers \et\ 2003}}
\def\Bond02{{Bond \et\ 2003}}
\def\Readhead03{{Readhead \et\ 2003}}

\def\metropolis53{{Metropolis \et\ 1953}}

\def\kowsowsky02{{Kosowsky \et\ 2002}}
\def\DASIpol02{{Leitch \et 2002, Kovac \et\ 2002}}
\def\BIMA02{{Dawson \et\ 2002}}
\def\bc01{{Bond \& Crittenden 2001}}
\def\bh95{{Bond 1996}}

\def\Freedman01{{Freedman \et\ 2001}}
\def\bbn03{{Kirkman \et\ 2003}}

\def\CKK03{{Chu, Kaplinghat \& Knox 2003}}
\def\VSA02{{Scott \et\ 2002}}
\def\VSAext02{{Grainge \et\ 2003}}
\def\BPS00top{{Bond, Pogosyan \&  Souradeep 2000}}
\def\archeops02{{Benoit \et\ 2003}}

\begin{document}

%%%%%%%%%%%%%%%%%%%%%%%

\begin{titlepage}

%%%%%%%%%%%%%%%%%%%%%%%%

\baselineskip 7ex
\mbox{}\vspace*{1.5ex}

\begin{center}
\Large {\bf The Cosmic Microwave Background and Inflation Parameters}
\end{center}

\vspace{2ex}

\baselineskip 5ex

\begin{center}
J. Richard Bond$^{1}$, Carlo Contaldi$^{1}$, Antony Lewis$^{1}$ and Dmitry
Pogosyan$^{2}$ \\
{\it 1. CIAR Cosmology \& Gravity 
Program, \\ Canadian Institute for Theoretical Astrophysics, Toronto,
Ontario, Canada \\ 
2. Physics Department, University of Alberta,
Edmonton, Alberta, Canada}
\end{center}

\vspace{7ex}

\begin{abstract}
We review the currrent cosmic parameter determinations of relevance to
inflation using the WMAP-1year, Boomerang, CBI, Acbar and other CMB
data. The basic steps in the pipelines which determine the bandpowers
from the raw data from which these estimations are made are
summarized. We forecast how the precision is likely to improve with
more years of WMAP in combination with future ground-based experiments
and with Planck. We address whether the current data indicates strong
breaking from uniform acceleration through the relatively small region
of the inflaton potential that the CMB probes, manifest in the
much-discussed running spectral index or in even more radical
braking/breaking scenarios. Although some weak ``anomalies'' appear in
the current data, the statistical case is not there. However increased
precision, at the high multipole end and with polarization
measurements, will significantly curtail current freedom.
\end{abstract}
\vspace{6ex}

\end{titlepage}

to appear in:  Int. J. Theor. Phys. 2004, ed. E. Verdaguer, "Peyresq Physics 8",
"The Early Universe: Confronting theory with observations" (June
21-27, 2003)

%\tableofcontents

\section{\hspace{-2.5ex}. Introduction to CMB Power Spectra}\label{sec:introduc}

\subsection{Overview}

The three Peyresq lectures covered CMB theory and analysis.  The main
content and most of the relevant references are given in Bond,
Contaldi and Pogosyan (2003, hereafter BCP).  Although a summary of
that material will be given here, in this paper we will emphasize the
impact on inflation phenomenology of the experiments WMAP, Boomerang,
CBI and Acbar, in conjunction with DASI, VSA, Archeops, Maxima, TOCO,
earlier CMB experiments and with large scale structure (LSS)
observations. We will also remind the reader of the great success now
of CMB determinations of the material content of the universe,
including dark energy and dark matter.  A detailed treatment of LSS
and the relation to the Sunyaev-Zeldovich (SZ) effect, the
cluster-dominated upscattering of CMB photons from hot electrons in
the cosmic web, was covered in the third lecture, and in Bond \etal
(2004), Readhead \etal (2004); see also Bond and Crittenden (1999).
At Peyresq, Page looked to the future with WMAP and the Atacama
Cosmology Telescope (ACT) and de Bernardis covered Boomerang and
beyond, concentrating on polarization. We will also explore here how
such future CMB observations may help to further discriminate among
inflation models. 

The theory of Gaussian primary anisotropies, those arising from linear
physics operating in the early Universe, is in good shape. For the
current data, speedy codes efficiently using past-history integrations
such as CMBfast and CAMB are quite adequate, and have been
``validated'' with codes solving hierarchies of multipole equations. As
precision in the data improves, fresh looks at all aspects of the
accuracy are worthwhile, and are being done.

For secondary anisotropies, those arising once nonlinearity develops,
the computational state of the art continues to need much further
effort.  This includes the important component which rears it head at
small angular scales associated with the SZ effect. The statistically
inhomogeneous Galactic foregrounds offer even more of a theoretical
challenge, and this has only partly been addressed. The direct
interface with observations for these many non-Gaussian signals is
much more complex than for Gaussian primary anisotropies. Because all
the signals are superimposed, the separation of the primary,
secondary, foreground and extragalactic components inevitably
complicates the move from multifrequency CMB data to determination of
cosmic parameters from ``cleaned'' primary CMB power spectra.

Major efforts by many groups around the world have been put into
developing the statistical pipelines which process and clean the raw
CMB data to allow efficient and accurate confrontation with the
theory. Even if primary power spectra are the primary target to be
extracted from the data, the case so far, the sophistication level
required is high: processing timestreams or interferometer
visibilities, making maps in position or ``momentum'' space,
filtering, cleaning, separating component signals, compressing, always
with new tools to explore systematic effects and anomalies that
inevitably appear.  The step from raw data to primary power spectrum
is enormous, from power spectrum to parameters small. Most of the
developments were driven by the compelling necessity of the CMB teams,
consisting of theorists/analysts as well as experimentalists, to
extract accurate science from beautiful large datasets such as
Boomerang and CBI, yet remain within computational
feasibility. Polarization analysis is receiving much further attention
now, and in spite of the many algorithmic advances in the analysis
pipeline in recent years, many more are needed for the rosy forecasts
of high precision described here to be realized.

\subsection{Grand Unified Spectra of CMB data}

\begin{figure}
\centerline{\hspace{0.1in}
\epsfxsize=6.0in\epsfbox{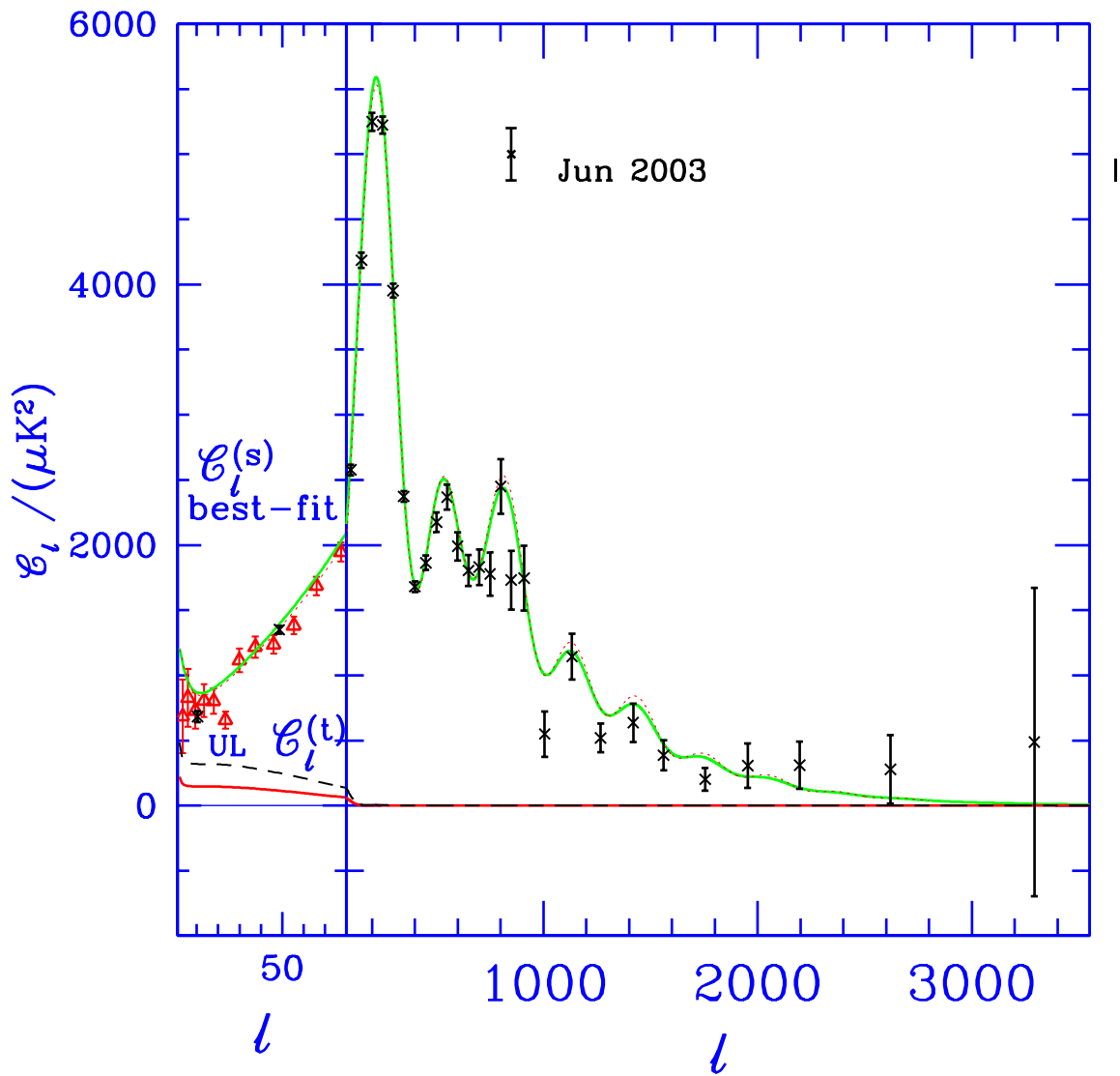} }
\caption{An optimal Grand Unified Spectrum for the post-WMAP Jun03
data is shown as crosses.  This GUS is a maximum-likelihood
determination of the power in 26 (top-hat) bands, with calibration and
beam uncertainties of the various experiments fully taken into
account. A best-fit inflation-based uniform-acceleration $\Lambda$CDM
${\cal C}_\ell$ spectrum to this Jun03 data is shown as solid green.
The parameters are $\{\Omega_{\rm {tot}}$, $\Omega_\Lambda$, $\Omega_b
h^2$, $\Omega_{\rm cdm} h^2$, $n_s$, $\tau_C, t_0, h, \sigma_8\}$ =
$\{1.0, 0.712, 0.0224, 0.118, 0.957, 0.108, 13.7, 0.698, 0.84 \}$. The
WMAP1 data optimally compressed on to 49 bandpowers is also shown for
low $\ell$ as triangles in the lower $\ell$ part of the figure, to
highlight the two low-$\ell$ ``anomalies'', at $\ell$ of a few and at
$\ell \sim 20$.  The best-fit ${\cal C}_\ell$ would fit better with a
slight downward tilt below $\ell \sim 30$ and beyond $\ell \sim 500$,
which a scale-dependent $n_s(k)$ could do (\S~\ref{sec:params}).
Peak and dip locations derived from this optimal spectrum are shown
in Fig.~\ref{fig:CLpkdip}. 
}
\label{fig:CLoptjun03}
\end{figure}

Fig.~\ref{fig:CLoptjun03} shows a ``Grand Unified Spectrum'' ${\cal
C}_\ell$ as a function of multipole $\ell$ compressed on to a small
number of bandpowers, derived from a June 2003 compilation of CMB
data. A best-fit inflation-based uniform-acceleration model to that
data is also shown. In place of determining ``cosmological
parameters'' from the observed bandpowers for various experiments, the
parameters here are in fact power amplitudes in multipole bands chosen
by the analyst. Other parameters characterizing experimental
uncertainties in calibration and beams are also determined
simultaneously. In determining errors on the cosmological
parameters and/or the GUS bandpowers, the other variables are
marginalized (probability distributions are integrated).

To see the remarkable evolution that has occurred in ${\cal C}_\ell$
in just a few years we refer the reader to Fig.~1 of BCP. This shows a
sequence of 4 GUS derived for the data extant in Jan00, Jan02, Jun02
and Jan03. (For experimental acronyms and much more detailed
discussion of how the results from the different experiments were used
see BCP.) The GUS are in excellent agreement with each other and with
the first-year WMAP results, hereafter WMAP1. All pre-WMAP1 CMB
results relied heavily on COBE's DMR results, which anchored the low
$\ell \lta 30 $ multipoles.  Fortunately WMAP1 spectacularly confirmed
DMR.  Jan00 included TOCO and the Boomerang North America test flight,
as well as 19 previous experiments, including upper limits. Jan02
included the Boomerang long duration balloon flight, DASI and Maxima
as well. Jun02 had CBI and VSA added as well. By Jan03, improved
Boomerang results were added to preliminary 2-year CBI results,
extended VSA results, Acbar and Archeops.

By Mar03, the WMAP1 data were incorporated, and recalibrated CBI 2
year and VSA results were included. The recalibration was tiny but the
errors on the calibration decreased by a factor of 2. The latter was
done off WMAP1, using observations of Jupiter. In GUS methods the
relative calibrations and their errors come out as a byproduct. BCP
showed these were in excellent agreement with cross-comparisons made
by Eric Hivon between WMAP1 and Boomerang maps, and with the CBI
calibration using WMAP1.  A Jun03 compilation utilized GUS-based
recalibrations of Boomerang and of ACBAR. The Jun03 GUS shown in
Fig.~\ref{fig:CLoptjun03} differs only slightly from the Mar03 GUS
used in BCP. This is basically because experimental calibrations and
beam sizes are internally determined and marginalized over in making
the GUS: if the method is correct it will give robust results -- and
it does. The CBI 2 year data has now been released (Readhead \etal
2004). The Jun03 GUS in Fig.~\ref{fig:CLoptjun03} does not include the
new VSA data (Dickinson \etal 2004), but with all the other data it does
not change the spectra or cosmological parameter determinations much.

WMAP1 dominates the $\ell \lta 600$ bands. Unless explicit joint
analyses are done of experiments with significant overlap with WMAP's
all-sky coverage, the bandpowers for such experiments should be
dropped from the GUS and parameter determinations. Thus COBE and
Archeops were not included in the Mar03 and Jun03 compilations.

\subsection{Peaks, Dips and Damping}

\begin{figure}
\centerline{\hspace{0.1in}
\epsfxsize=6.0in\epsfbox{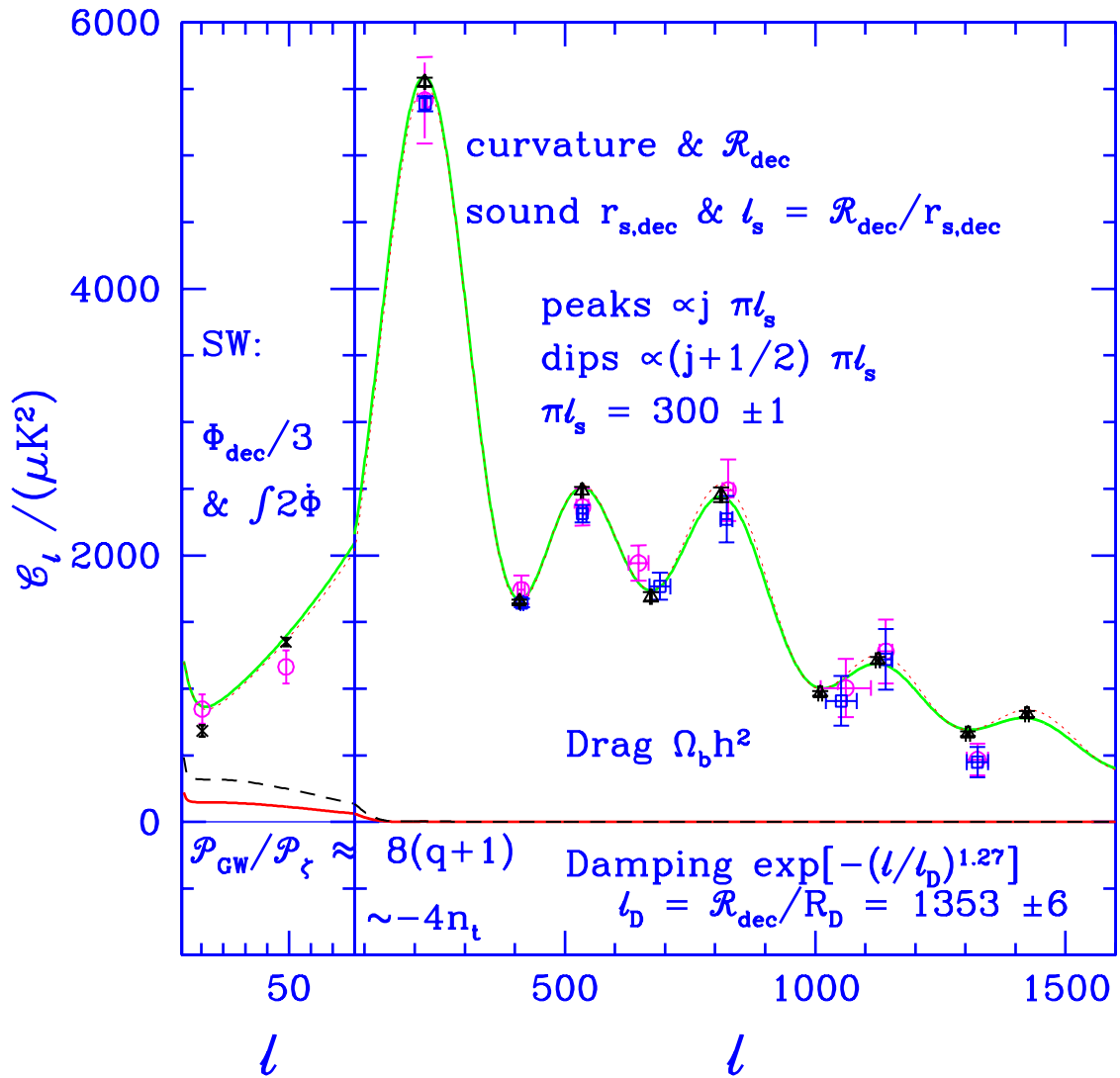} }
\caption{Various pillars which determine the inflation-based
uniform-acceleration ${\cal C}_\ell$ spectrum shown that best-fits the
Jun03 data are highlighted. The low $\ell$ part of the figure repeats
as crosses the two Jun03 bands of Fig.~\ref{fig:CLoptjun03} and
contrasts these with the equivalent two bands for the pre-WMAP1 Jan03
data, which includes COBE and Archeops. The low $\ell$ shape arises
both from the climb of the photons through the potential well at
decoupling and the propagation of the photons through a time varying
gravitational potential along the line of sight. Also shown at the
bottom is the expected tensor component for uniform acceleration with
tensor tilt $n_t$ = $n_s-1$ = $-0.043$, and above it the Spergel \etal (2003) 95\%
CL limit, corresponding to ${\cal P}_{GW} / {\cal P}_{\zeta} < 0.36$.
The higher $\ell$ part of the figure shows the peak/dip locations
$\ell_{pk/dip,j}$ and heights ${\cal C}_{pk/dip,j}$, as determined
from the BCP maximum likelihood sliding-band approach. The points with
slightly larger errors are for the pre-WMAP1 Jan03 data and the heavier
are for the post-WMAP1 Jun03 data. The triangles show the values
obtained by ensemble-averaging peak/dip locations and heights over the
large ${\cal C}_\ell$-database used in BCP. (Only a weak prior was
applied, which allows large movement of peak locations associated with
the geometry, hence is preferable to more restrictive priors for this
application. Note how well these statistically-averaged peaks and dips
match those of the best-fit model.)  }
\label{fig:CLpkdip}
\end{figure}

Fig.~\ref{fig:CLpkdip} illustrates some of the ``pillars'' that we
were looking for in the TT data to verify the paradigm, exemplified by
the best-fit power-law model. (1) The effects of a large scale
gravitational potential at low multipoles, manifested by a
``Sachs-Wolfe'' plateau; note the integrated SW upturn associated with
$\Omega_\Lambda$ driving a time-varying gravitational potential at low
$\ell$ and the upturn at higher $\ell$ associated with photon
compressions and rarefactions.

At higher $\ell$ the next pillars are (2) the pattern of acoustic
peaks and dips and (3) the damping tail. These are governed by the
comoving sound speed $r_s = 146 \pm 3 \mpc$ and damping length $R_D=10.3
\pm 0.21 \mpc$ at photon decoupling, and are scaled with the
angular-diameter distance relation ${\cal R}_{dec}$ to decoupling to
get the associated $\ell_s$ and $\ell_D$ values which define the
peak/dip and damping structure.  

Fig.~\ref{fig:CLpkdip} shows the peak/dip locations and amplitudes and
their 1-sigma error bars, as determined in BCP, contrasting Jan03 with
Jun03.  The exercise in BCP was to do this directly from the
relatively broad-band GUS, using a model that slides a sequence of
bands across the data sets. In spite of the coarseness of the bands, the
peak/dip results are highly accurate as long as all band-to-band
correlations are taken into account. These peak and dip parameters
have also been determined for individual experiments such as TOCO,
Boomerang, CBI, Archeops, and of course for WMAP1, the latter
described in Page \etal (2003). Values are given in BCP, Table~2.

There is good evidence from WMAP1, Boomerang and other CMB analyses
that the statistics of the primary anisotropies are predominantly (4)
Gaussian, \ie\ have maximal randomness for a given power
spectrum. Finding (5) secondary anisotropies associated with nonlinear
phenomena, due to the SZ thermal and kinetic effects, inhomogeneous
reionization, weak lensing, \etc\ is expected. The anomalous extra
power at high $\ell$ over the best-fit primary model evident in
Fig.~\ref{fig:CLoptjun03} arises from combining CBI and Acbar
data. Assuming an SZ spectrum makes the case somewhat stronger,
and suggests this pillar may have been seen (see Fig.~\ref{fig:pzeta}
and Bond \etal 2004, Goldstein \etal 2003. Readhead \etal 2004).

\begin{figure}
\centerline{\hspace{0.1in}
\epsfxsize=6.0in\epsfbox{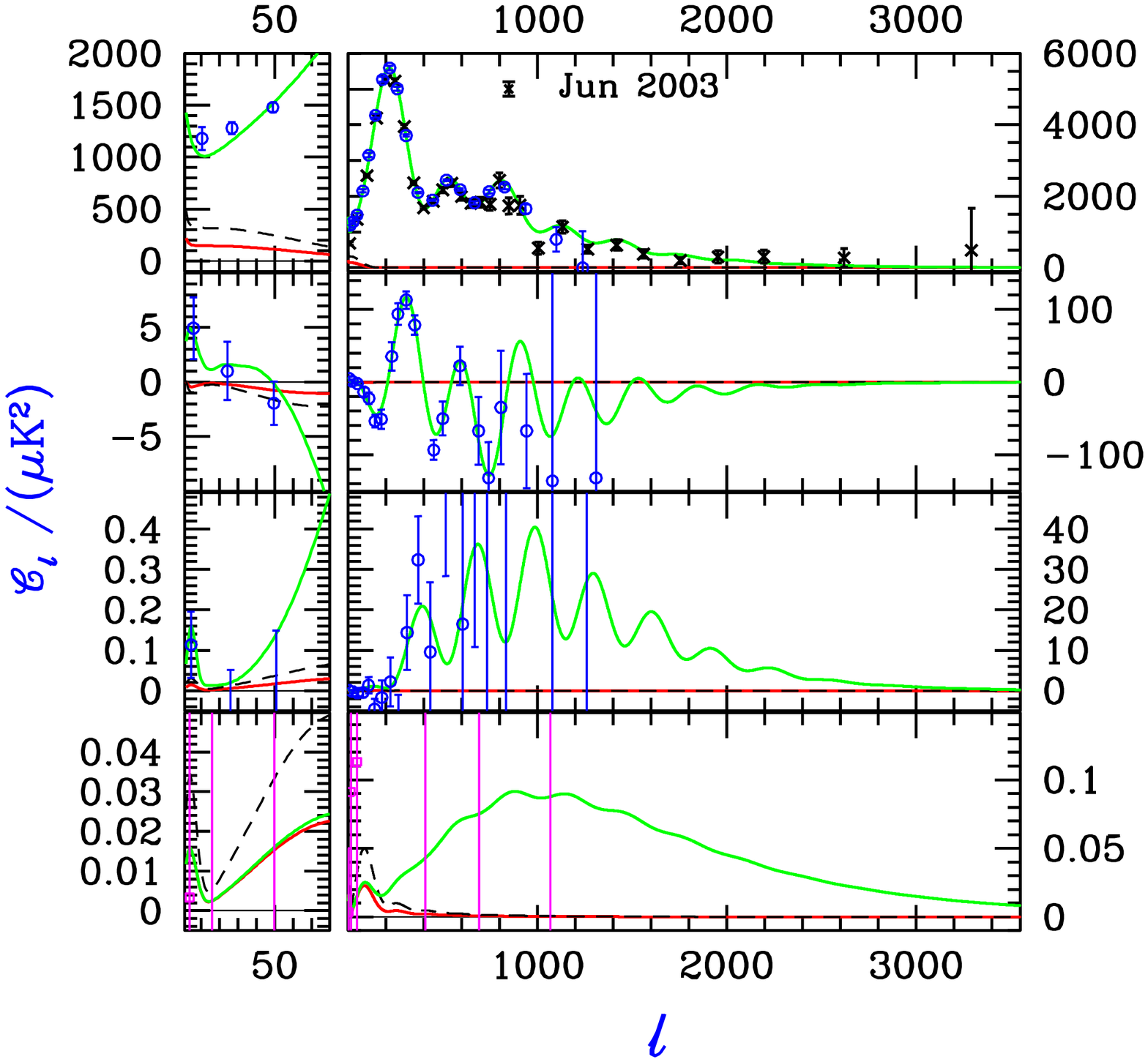} }
\caption{Idealized forecasts for detections of (top to bottom) TT, TE,
  EE and BB bandpowers by WMAP with four years of 94 GHz channel
  data. This is a conservative estimate given the other WMAP
  frequencies, but idealized in the sense that foregrounds and other
  experimental complications are ignored. Simulations use the best-fit
  Jun03 model shown. The Jun03 GUS are the crosses in the TT
  panel. Note the very different scales on the low and high $\ell$
  panels of the figure. As well as the target scalar ${\cal
  C}_\ell^{(s)}$, the tensor ${\cal C}_\ell^{(t)}$ contributing at the
  level predicted if $n_t = n_s-1$ is also shown. The tensor shape is
  repeated with the amplitude corresponding to the current WMAP1 upper
  limit (dashed) on the tensor to scalar ratio.  Primary scalar
  perturbations have no B-mode in linear perturbation theory, but they
  are induced by lensing (Hu 2000). These and the unknown foregrounds
  present severe challenges for confirmation of the gravity-wave
  induced B-mode. The B-mode bandpower spacings differ from the
  spacings used for the upper panels. }
\label{fig:pol4_wmap4}
\end{figure}

\begin{figure}
\centerline{\hspace{0.1in}
\epsfxsize=6.0in\epsfbox{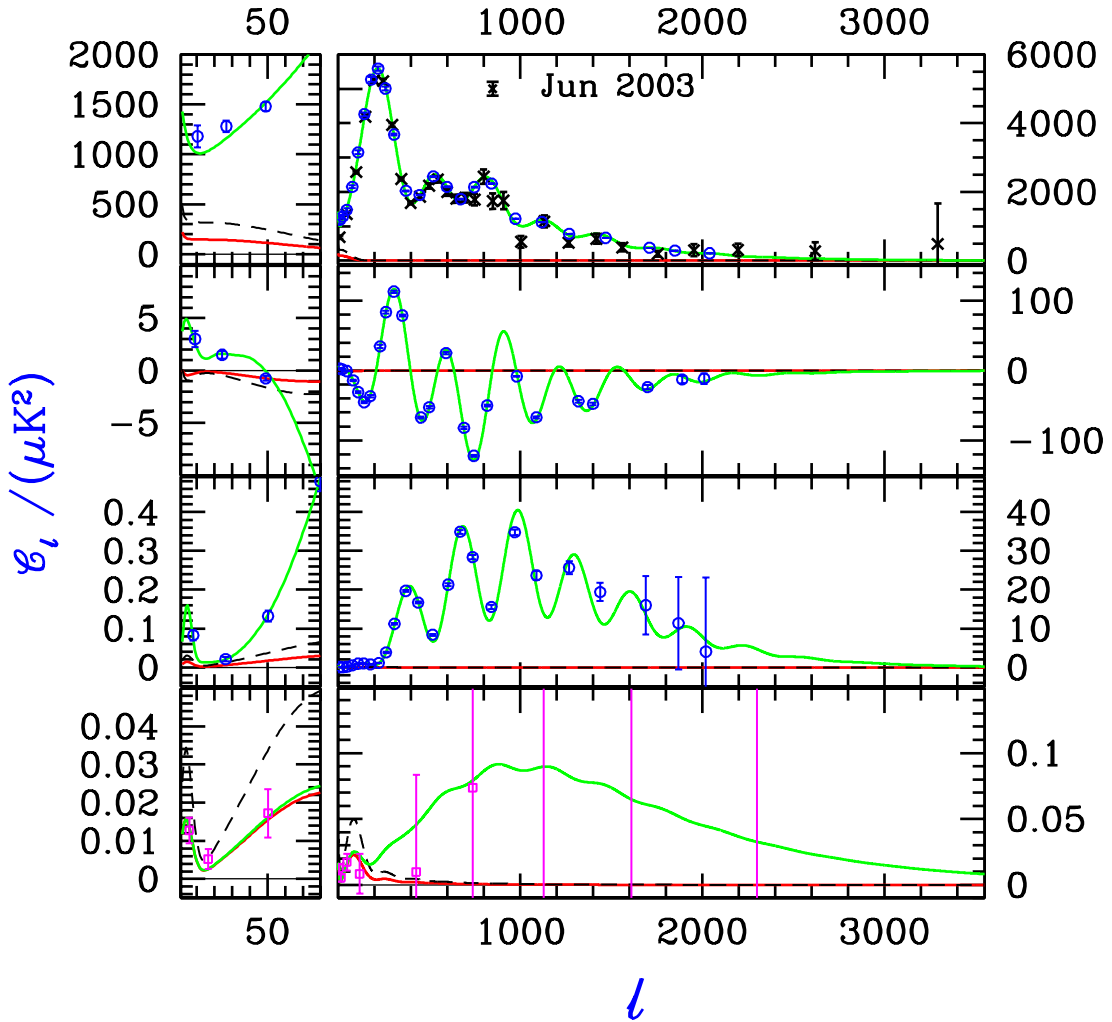} }
\caption{Forecast for how well Planck can do with just its 150 GHz
  channels for one year of data, quite a conservative estimate. Even
  so, note the anticipated detection level of EE and TE at low $\ell$,
  sharpening the $\tau_C$ determination, and the possibility of a
  statistically significant direct tensor-induced B-mode detection.  Although
  ground-based experiments at high resolution should have a huge
  pre-Planck impact (\eg\ Fig.~\ref{fig:pol4_wmap4_SPT1000}), the
  all-sky nature of Planck, its large set of polarization-sensitive
  frequencies and likely longer observing time than that assumed here will
  make it extremely powerful to sort out the many signals that
  complicate the ``primary'' quest.  }
\label{fig:pol4_planck1}
\end{figure}

\begin{figure}
\centerline{\hspace{0.1in}
\epsfxsize=6.0in\epsfbox{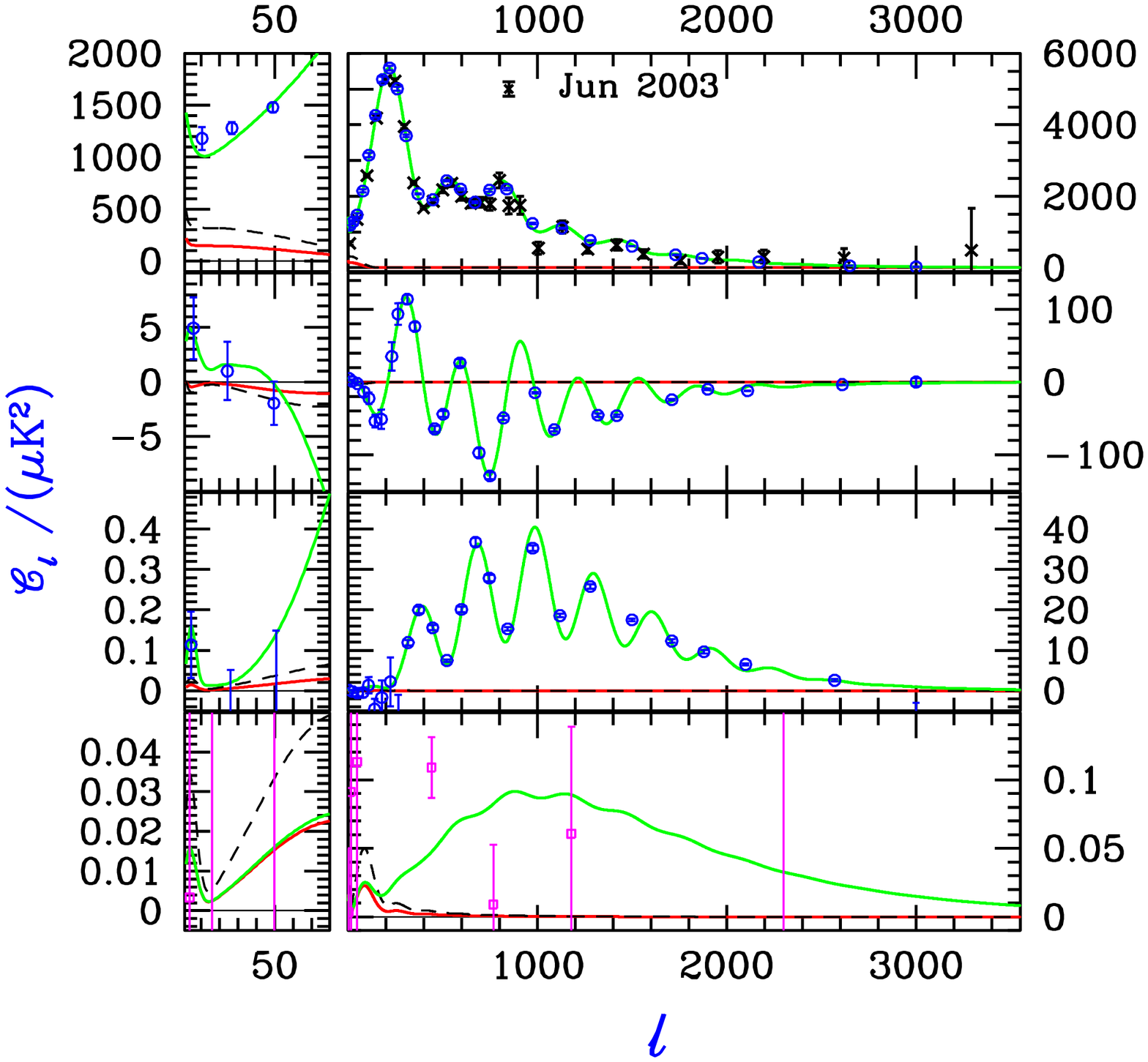} }
\caption{Forecast for how well a high-resolution ground-based
  polarization-sensitive large-array experiment like ACT or SPT can
  do, in conjunction with WMAP4. The specific assumptions used are
  given in the caption of Fig.~\ref{fig:fcastrun} and are conservative
  over what might be achievable from the ground using bolometer
  arrays. There is a trade off of sky coverage and noise. 
  ($f_{sky}=0.024$ here, with larger and shallower often improving
  cosmic parameter determinations, and smaller and deeper making the
  lens-induced B-modes potentially detectable.) Apart from the current
  DASI, WMAP, Boomerang and CBI data already in to flesh out the EE
  spectrum, a number of other experiments are planned. These include
  some in the very near future, \eg QUaD and BICEP. Forecasts for the
  proposed QUIET with HEMT arrays look as promising as those for
  polarized ACT/SPT. Given the expected signal levels, all ground and
  space CMB information available will be needed and used to get clean
  primary polarization results. }
\label{fig:pol4_wmap4_SPT1000}
\end{figure}

(6) Polarization must be there, with forecasted ${\cal
  C}_\ell^{(EE)}$-patterns of peaks and dips intimately related to,
though with different phases than, those for TT. As well there is a
specific peak/dip pattern in the TE cross-correlation of the E-mode
with the total intensity predicted. The current status of polarization
is a broad-band EE detection consistent with inflation by DASI and
of course the remarkable TE cross correlation of WMAP1. More EE detections
are expected soon from WMAP2, CBI and Boomerang, and there are many
planned experiments that should exquisitely determine the EE and TE
spectra. Figs.~\ref{fig:pol4_wmap4}, \ref{fig:pol4_planck1} and
\ref{fig:pol4_wmap4_SPT1000} show the TE and EE spectra for the
best-fit model and forecasts of polarization detections by WMAP4, by
Planck1, and by a fiducial ground-based large-array
polarization-sensitive telescope.  

A future goal for CMB researchers is to find (7) the anisotropies
induced by gravity wave quantum noise. Not all inflation models
predict this, but the well known simple relation between tensor tilt
$n_t$, the deceleration parameter $q$ and the tensor-to-scalar power
ratio shown in Fig.~\ref{fig:CLpkdip} suggests there may be a strong
enough signal to detect. For the Jun03 best-fit model shown with
scalar index $n_s=0.957$, a uniform acceleration model yields a
tensor-to-scalar contribution of 0.17 {\it cf.} the Spergel \etal
(2003) upper limit of 0.36 at the 95\% CL. A holy grail for the
subject is to detect the B-mode of polarization at low $\ell$. For
these best-fit parameters, it may be do-able with the Planck satellite
as the lower panel of Fig.~\ref{fig:pol4_planck1} illustrates. 
A nice figure summarizing EE and BB bandpower forecasts for various
experiments such as QUaD and BICEP in comparison with Planck is given in
Hivon \& Kamionkowski (2002). Our forecasts for these are in accord. 

The very tiny B-mode signal predicted and the unknown nature of the
polarized foregrounds would make this a very hard task indeed, one
happily defining a long term future for CMB research as the community
plans a future NASA CMBPol satellite as the next step in
space after Planck.

\section{\hspace{-2.5ex}. Cosmic Parameters \& Inflation Issues}\label{sec:params}

\subsection{The Conventional Minimal Parameters}

To the well known minimal inflation-motivated set, $\{\Omega_b{\rm
h}^{2},\Omega_{cdm}{\rm h}^{2} ,\Omega_{k}, \Omega_\Lambda,n_s,\tau_C,
A_s\}$, defining the allowed ${\cal C}_\ell$ shapes, other parameters
are sequentially added to probe more complex models of inflation or of
matter content. ($\Omega_j$ is the density parameter $\rho_{j}$
relative to the critical density $\rho_{cr} = 10.5 \, {\rm h}^{2} \,
{\rm kev} \cc$, where ${\rm h}$ is the Hubble parameter in units of
100 $\kms \, {\rm Mpc}^{-1}$, so $\Omega_b {\rm h}^{2}$ and
$\Omega_{cdm}{\rm h}^{2}$ are proportional to the physically more
relevant baryon and cold dark matter densities.)  The vacuum energy is
$\Omega_{\Lambda}$ and the mean curvature energy for our Hubble patch
is $\Omega_k$, in terms of which the total energy content is
$\Omega_{tot}=1-\Omega_k$. Sample content extras include a subdominant
light neutrino component, enhanced relativistic particles
$\Omega_{er}{\rm h}^{2}$, \eg as products of decay of massive
particles, adding dynamics to $\Omega_\Lambda$, \eg through a
parameterized time-dependent pressure-to-density ratio,
$w_\Lambda (t)$.

Astrophysical complications associated with the ``late time'' impact
of reionization of the Universe, presumably by the injection of energy
from the first stars, are encoded in a single parameter $\tau_C$, the
depth to Compton scattering when the Universe reionized. $\tau_C$
could have a complex temporal and spatial structure, and, although the
CMB primary anisotropies are not that sensitive to the details,
finding such signatures in the data is a target of planned
high resolution experiments.

\subsection{The Inflation Parameters}

Inflation fluctuations are assumed to have a Gaussian statistical
distribution, fully specified by a power spectrum of curvature
perturbations ${\cal P}_{\varphi_{com}}(k)$. The minimal set has only
two parameters, the overall initial power spectrum amplitude $A_s
\equiv {\cal P}_{\varphi_{com}}(k_n) $, evaluated at a normalization
wavenumber $k_n$, and a single spectral scalar index $n_s(k)=n_s(k_n)$.

There is of course a vast literature on perturbation theory as applied
to inflation (\eg Bond 1994, 1996 for the approach described
here). Basic variables are the inflaton field $\delta \phi_{inf}$;
other scalar field degrees of freedom $\delta \phi_{is}$ which can
induce scalar isocurvature perturbations; two gravitational wave
polarization modes $h_{+},h_\times$.  One can encode scalar metric
perturbations and their variations through the inhomogeneous scale
factor $a(\bx ,t)$, Hubble parameter $H(\bx ,t)$ and deceleration
parameter $ q(\bx , t) \equiv -d \ln Ha / d\ln a$.  For example, the
scalar 3-curvature is $-4 \nabla^2 \ln a$. Inflation ends when $q$
passes from negative to positive.  

Certain time hypersurfaces upon which to measure the
perturbations simplify things quite a bit. Sample choices are
$\phi_{inf}$, $\ln a$, $\ln H$, $\ln (Ha)$ and conformal time $\tau =
\int d\ln a /(Ha) $. The power spectrum $ {\cal P}_{\ln
a\vert_{H_*}}(k) $ of scale factor fluctuations evaluated on time
hypersurfaces on which the Hubble parameter $H$ is constant becomes
time-invariant for wavenumbers $k/Ha \ll 1$ (outside of the
``horizon''), if there is just one dynamically important scalar field,
and remains so until fluctuations regain causal contact. If the
universe has no net mean curvature, $\varphi_{com} =\delta \ln
a\vert_{H_*} $, measuring the curvature on comoving
hypersurfaces. Another variable used extensively is $\zeta $, which
reduces to $\varphi_{com} $ if $k/Ha \ll 1$, hence we sometimes refer
to ${\cal P}_\zeta$ in the figures.

The gravity wave power ${\cal P}_{GW}(k) \equiv (k^3
/2\pi^2)<h_+^2 + h_\times^2 >$ = $(k^3 /2\pi^2)<h_{ij}h^{ij}>/2$ used
here and in Bond (1994, 1996) is defined to be consistent with the
conventions of GW detection research, with GW mode functions being the
usual $ h_\times = h_{12} = h_{21}, \ h_{+}=(h_{11}-h_{22})/2$. Most
people in inflation define a ${\cal P}_{h}(k) = 2 {\cal P}_{GW}(k)$,
which is what $A_t \equiv {\cal P}_{h}(k_n) $ is defined in terms of.  

Quantum fluctuations in gravity waves must occur during inflation.
The only question is how ${\cal P}_{GW}(k)$ compares with ${\cal
P}_{\varphi_{com}}(k) $. Thus the minimal $A_s, \, n_s$ should at
the very least be augmented by an $A_t$ and $n_t$. As well subdominant
isocurvature components can be added, $A_{is}$ and $n_{is}$. 

For all of these cases, ``radically broken scale invariance'' may
prevail, in which the spectral index functions
\begin{eqnarray}
n_s(k) -1 &\equiv& d\ln {\cal P}_{\varphi_{com}}(k)/d\ln k = 2(1 +q^{-1}) +
q^{-1} q^{\prime}/(1+q) +\ 2C_s \, , \label{eq:ns} \\ 
q^{\prime} &=&  dq/d\ln a = -q dq/d\ln (Ha) \, ,  \nonumber \\
n_t(k) &\equiv&  d\ln {\cal P}_{GW}(k)/d\ln k = 2(1 +q^{-1}) \ +2C_t \, ,  \label{eq:nt}
%n_{is}(k) &\equiv& d\ln {\cal P}_{is}(k)/d\ln k = 2(1 +q^{-1}) \ +2C_{is} \, , \label{eq:nis}
\end{eqnarray}
can be relatively arbitrary functions of spatial wavenumber rather
than constants. Here quantities such as $q$ and $q^\prime$ 
are evaluated at the ``time'' $ Ha =k$. The formula is motivated by the
stochastic treatment of inflation fluctuations in the
Hamilton Jacobi framework, in which  quantum noise at the
Hawking temperature $H /2\pi$  radiates from short distances across the
decreasing $(Ha)^{-1}$ boundary into a long wavelength background
field. The post-inflation power spectra
are parameterized by
\begin{eqnarray}
{\cal P}_{\ln a\vert_{H_*}}
&=& {1\over {q+1}} \, {{4\pi} \over \mpl^2} \, 
(H /2\pi )^2  \, e^{2u_s} \, , \quad  {\cal P}_{GW} = 8 \, {{4\pi} \over \mpl^2} \, 
(H /2\pi )^2  \, \, e^{2u_t}  \ .   \label{eq:PTPS} 
\end{eqnarray}
Of course the utility of these expressions depends upon the correction
factors $C_{s, t}$, which are derived from the related
$u_{s,t}$. Analytical forms for special cases can be derived, \eg for
uniform acceleration, and these show the $C_{s, t}$ are typically
small.\footnote{The stochastic inflation technique uses ``the $H/(2\pi
)$ at $k=Ha$ WKB approximation'', writes eq.(\ref{eq:PTPS}) as a
function of $H$, $q$ and derivatives, and takes a logarithmic
derivative {\it wrt} $Ha$ in place of $k$. Analytical corrections
invariably involve Hankel functions and their asymptotic
expansions. No slow roll ($(1+q)\approx 0$) restrictions are needed in
these approaches. In the HJ formulation, $H(\phi )$ and $q(\phi)$ are
treated as functions of the inflaton field, and satisfy the `reduced
Hamilton-Jacobi equation' relating $H(\phi)$ to the potential
$V(\phi)$: $H^2 = H_{SR}^2/( 1-(q+1)/3)$, where $H_{SR}^2 \equiv {8
\pi V\over 3 \mpl^2}$ and $(1+q) = {\mpl^2 \over 4 \pi} \,\Big[
{\partial \ln H \over \partial \phi}\Big]^2$. The extra piece in
$n_s-n_t $ is $q^{-1} q^{\prime}/(1+q)$ = $- q^{-1} {\mpl^2 \over 4
\pi} \, {\partial^2 \ln H \over \partial \phi^2}$, responsible for
deviations in the two indices that can be significant near $(1+q)
\approx 0$.}  Much has been written on this subject. See, for example,
Bond (1994), Lidsey \etal (1997), Wang \etal (1997) Schwarz \etal
(2001), Kinney (2002), Habib \etal (2002), Martin and Schwarz (2003),
Peiris \etal (2003), Leach and Liddle (2003), and references therein.
The accurate path to the spectral indices is to take logarithmic
derivatives of full numerical calculations to get the $C_{s, t}$. One
can certainly invent cases in which the $C_{s, t}$ are not
small. However, provided $q$ does not change too rapidly it is
reasonable to use these formulas as guides. They show that tilt mostly
depends upon how far the acceleration is below the {\it critical}
value of unity. For $q\approx -1$, a substantial scalar tilt can come
from the second term, yet no tensor tilt, as in natural inflation.

Deviations from the power law model are not just expected, they are
necessary, since $q(k)$ must have passed from negative to positive to
have created matter from the vacuum energy housed in the inflaton.
The simplest form of braking of the acceleration is the running index,
$n_s (k) = n_s(k_n) + [dn_s/d\ln k (k_n)]\, \ln (k/k_n) $.  Such a
form can be expressed in terms of coefficients $q_n$, $q_n^\prime$,
$q_n^{\prime\prime}$ in an expansion of $q(k)$.

{\it A priori} it may seem unlikely that a marked change in the
expansion rate or acceleration would just happen to be in the 3-decade
window of $k$-space accessible to our CMB observations, since it maps
into a relatively small patch on the inflaton potential surface
because of the $(1+q)$ suppression factor in ${\sqrt{4\pi} \over \mpl}
d\phi = \pm q^{-1} (1+q)^{1/2} d\ln (Ha)$. That is that $q_n^\prime$
would be small. However in $\phi$-space, this CMB window is not very
far from $\phi_{end}$ defining the acceleration/deceleration boundary,
hence the $q$ rise to zero must be reasonably rapid in $\phi$. Even
so, for most inflation models, the rapid change does indeed occur only
near the end, suggesting special physics might have to be built in to
accommodate large change earlier.  Rapid acceleration changes, if present,
would seem to be more likely a consequence of interaction with other
field degrees of freedom rather than a result of inflaton
self-interaction. Such hybrids involving two scalars interacting with
either simple polynomial potentials (with second, third and fourth
order terms), combinations of exponential potentials, and other simple
forms, have long been used to show that constructing power spectra
with mountains and valleys and even generating non-Gaussian 
fluctuations is possible in inflation. 

However even if more than one scalar field enters it is often possible
to consider an effective single inflaton self-interacting through an
effective single-inflaton potential over the observable scales. This
is because the fields first settle into gorges on the potential
surface, then follow the gorge downward towards the local minimum
along a single field degree of freedom, $\phi_\parallel$, to be
identified with the inflaton. The other degrees of freedom,
$\vec{\phi}_\perp$, are `isocurvature' degrees of freedom. Usually,
the faces rising up from the gorge will be sufficiently steep that the
inevitable quantum noise that excites motion up the walls quickly
falls back, leaving no usable isocurvature imprint, effectively making
those direections irrelevant (although curvature in the trough can lead
to complications in the kinetic energy piece of the inflaton degree of
freedom). The single-inflaton expressions in terms of $q$ would
prevail.  

To get observable response often involves invoking an instability,
with negative transverse components of the mass-squared matrix, $
{\partial^2 V /\partial\phi_i \partial \phi_j}$, leading to an opening
up of the gorge or its bifurcation.  Tuning the location of such a
structure to the window on the potential surface we can access may
seem to be unpalatably precise. This is perhaps mitigated by relating
it to a waterfall of sudden $q$ change to trigger reheating, 
termed hybrid inflation. Control of residual defects left from the
potential was always an issue, but such residual subdominant
components are worthwhile to hunt for in the data.  Although these
multiple field models would have significant deceleration occurring in
a waterfall phase, to have a spectrum with many sharp features
littering the CMB range parameterizing a complex braking pattern seems
very baroque indeed. (These multiple field situations are the ones
where the simple spectral index formulas in terms of $q$ are 
likely to have the largest corrections.)  CMB phenomenologists should
constrain such possibilities anyway.  We shall see that the prognosis
for constraining even such radical braking is reasonable with upcoming
CMB experiments.

The richness of inflation theory has expanded considerably with the
emphasis on higher dimensions, brane-ology and stringy cosmology. So
far to the extent there are predictions they fit within the basic
inflation phenomenology as applied to CMB analysis of the sort
described here. It would be nice if a smoking gun pointing to a
uniquely stringy culprit will be found theoretically, and in the data.

\section{\hspace{-2.5ex}. Parameter Estimations, Now and Future} \label{sec:param} 

\subsection{Parameters from the Jun03 Data} \label{sec:paramnow} 

\begin{figure}
\centerline{\hspace{0.1in}
\epsfxsize=6.0in\epsfbox{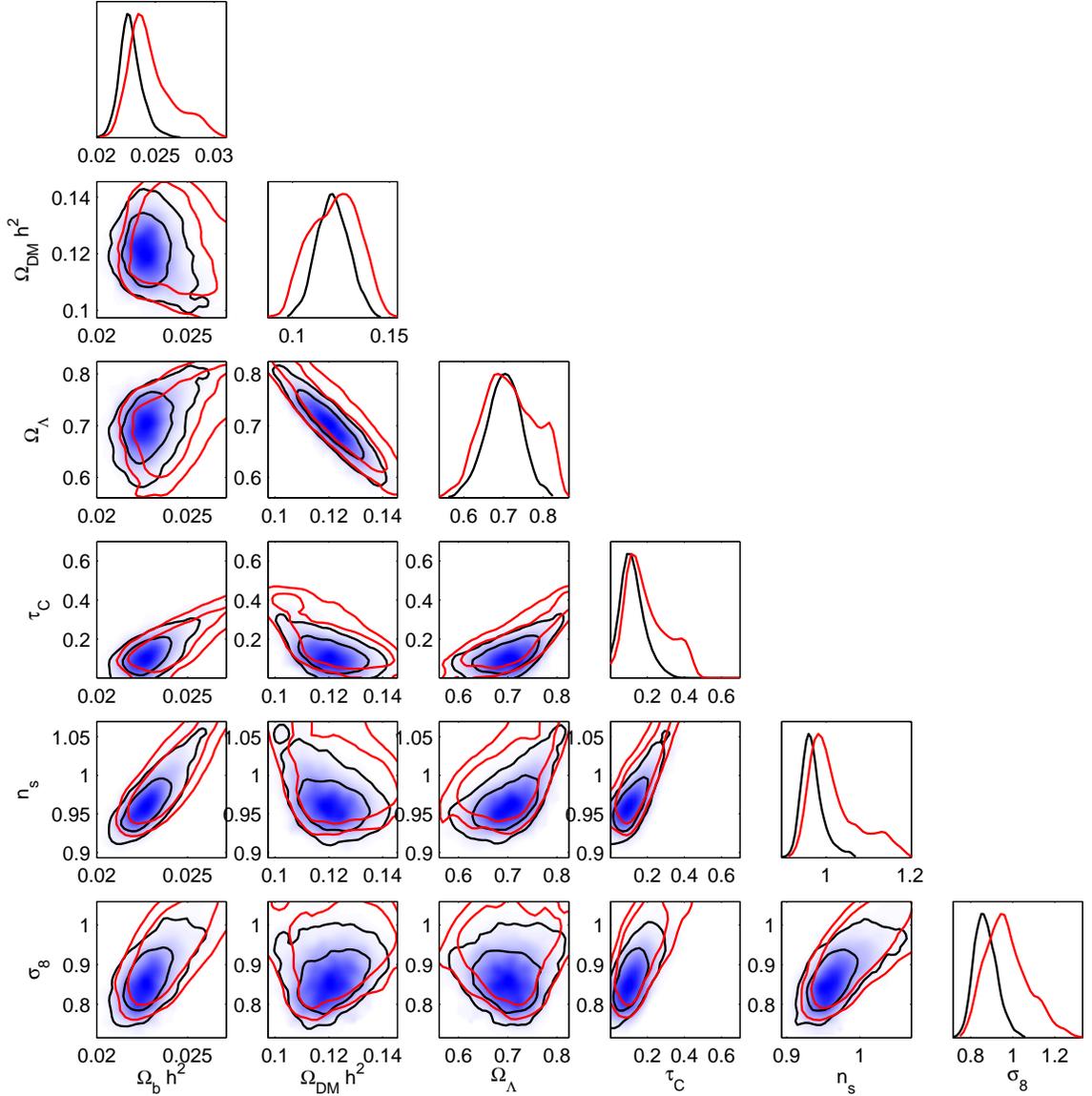} }
\caption{The state of parameter estimation using the Jun03 data
compilation is illustrated by the one and two sigma contour regions
when all but the two variables shown are marginalized.  The scalar
spectral index was not allowed to run. The outer contours are for
WMAP1 alone. The 1D probability distributions for each single variable
shown at right gives means and 1-sigma errors listed in
Table~\ref{tab:exptparamsMCMC}. This illustrates how the current data
to higher $\ell$, predominantly driven by Boomerang, CBI and Acbar,
sharpen the WMAP1 results by breaking partial degeneracies.  The
priors applied were $\Omega_k =0$ and the weak $h$-prior. In the
figures, the $\sigma_8$-dominated LSS prior was included as well. The
table shows the extent to which this sharpens parameter
determinations. }
\label{fig:cbiallLSS}
\end{figure}

Fig.~\ref{fig:cbiallLSS} shows visually the one and two sigma
constraints for the minimal inflation parameter set derived from the
Jun03 CMB dataset using the Monte Carlo Markov Chain method.  These
results were also reported in Readhead \etal (2004) and are very
similar to the Mar03 results given in BCP. BCP showed that the MCMC
results were also in good agreement with those obtained using fixed
${\cal C}_\ell$-grids. Table~\ref{tab:exptparamsMCMC} gives projected
one-sigma MCMC errors.  Priors applied to augment the CMB data with
other information range from weak ones to stronger constraints from
the expansion rate (HST-h), from cosmic acceleration from supernovae
(SN1) and from galaxy clustering, gravitational lensing and local
cluster abundance (LSS). We show results in the table for CMB+weak and
CMB+weak+LSS, with a flat $\Omega_{tot}=1$ prior also imposed.

\begin{figure}
\centerline{\hspace{0.1in}
\epsfxsize=6.0in\epsfbox{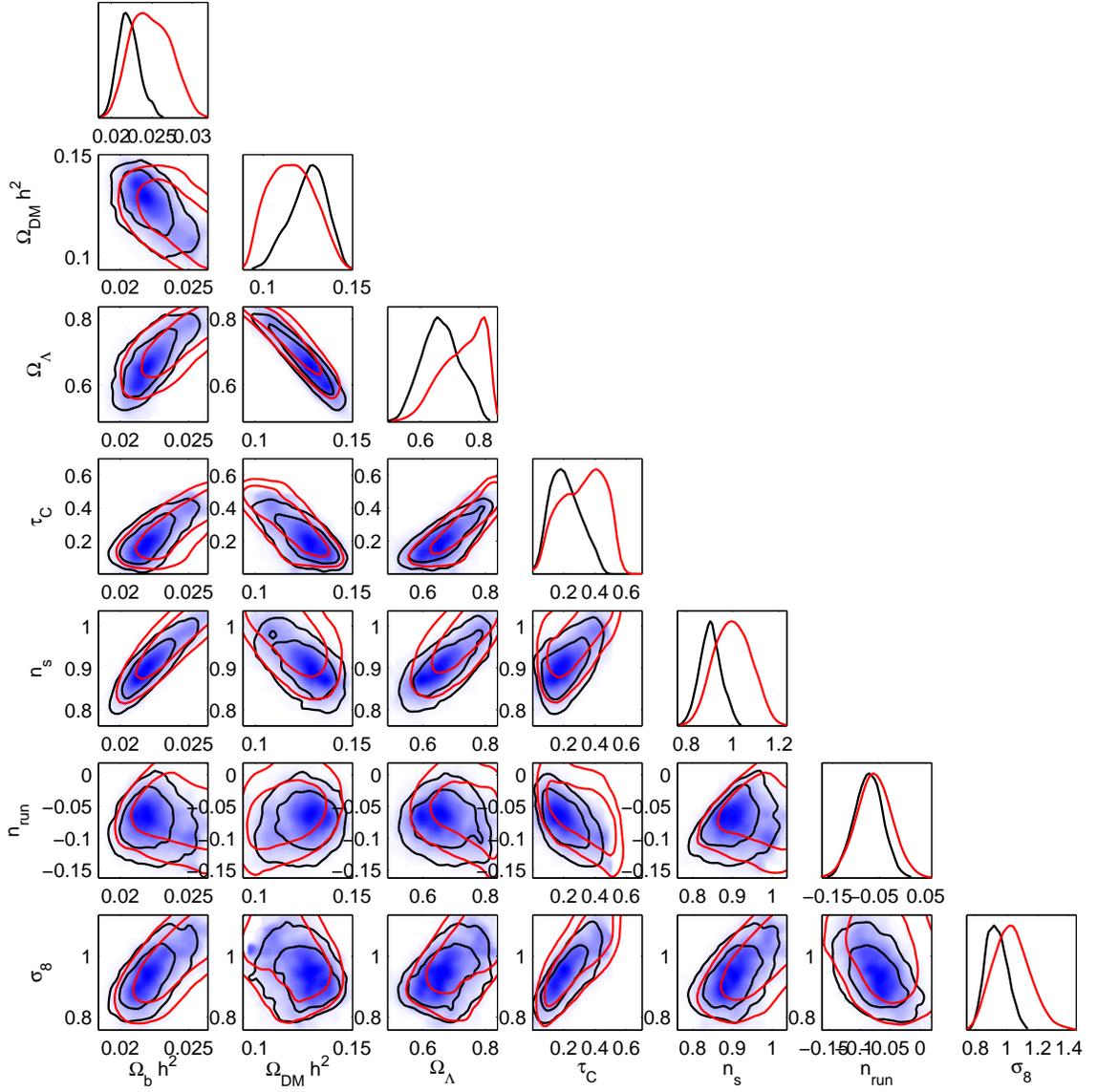} }
\caption{Similar to Fig.~\ref{fig:cbiallLSS}, except the scalar
spectral index is allowed to run. Note the $\sigma_8$ and $\tau_C$
shifts indicated here and in Table~\ref{tab:exptparamsMCMC}. }
\label{fig:cbialldnLSS}
\end{figure}

\begin{sidewaystable}
\label{mytable}
\caption{Sample cosmological parameter values and their 1-sigma errors
are shown, determined using MCMC methods after marginalizing over the
other cosmological parameters and, for real data, the various
experimental parameters. The first four columns show the state as of
Jan04. WMAP1 refers to the actual WMAP one year data, Jun03 is the
compilation including the WMAP1 data and recalibrated Boomerang, CBI,
VSA and Acbar data, along with DASI and Maxima data. Means and
standard deviations are given here. In all cases, the weak prior
($0.45 \le {\rm h} \le 0.9$, age $> 10$ Gyr) is applied, and $w_Q$ is
fixed at $-1$, the cosmological constant case.  The curvature parameter
$\Omega_k$ is fixed at zero. The LSS prior agrees with current weak
lensing data and agrees with most of the cluster determinations. A
weak redshift survey constraint is also imposed. The 2dF ``prior''
uses the stronger 2dF redshift survey results, but not a $\sigma_8$
prior. The primordial power spectrum index for scalar perturbations
obeys $n_s (k) = n_s(k_n) + [dn_s/d\ln k (k_n)]\,  (\ln k /k_n)$. For the first set
of numbers $dn_s/d\ln k (k_n) $ is set to zero (no
running of the spectral index). For the second set, it is allowed to
vary. The last two columns show how the errors are expected to improve
for WMAP with four years of data and Planck with one year of
data, for the Fig.~\ref{fig:mcmcrun} simulations. }
\label{tab:exptparamsMCMC}
\begin{center}
\begin{tabular}{|l|llll|ll|}
\hline
MCMC  & WMAP1 &  Jun03 & Jun03+LSS & Jun03+2dF & WMAP4 & Planck1 \\
\hline
flat+weak  &   &  &  &  &  &  \\
\hline 
$\Omega_b{\rm h}^2$ &
 $.0243 \pm .0018$ 
 & $.0228 \pm .00095$  & $.0229 \pm .00097$  & $.0228 \pm .00090$ & $.02271 \pm .00047$ & $.02256 \pm .00017$   \\
$\Omega_{cdm}{\rm h}^2$  
& $.123 \pm .018$ & $.118 \pm .011$   & $.121 \pm .009$   & $.120 \pm .007$ & $.1122 \pm .0039$  & $.1165 \pm .0016$   \\
$n_s$   & $1.01 \pm .054$ &
 $0.964 \pm .026$  & $0.967 \pm .027$ & $0.965 \pm .024$  &  $0.9704 \pm .0125$ & $0.9599 \pm .0045$  \\
$h$   & $0.721\pm 0.064$ & 
$0.705 \pm 0.041 $ & $0.697 \pm 0.036 $  & $0.696 \pm 0.028 $ & $0.729 \pm 0.021$ & $0.706 \pm 0.008$ \\
$\tau_C$ & $0.184 \pm .094$ &
$0.117 \pm 0.059$  & $0.123 \pm 0.061$  & $0.110 \pm 0.056$  &  $0.100  \pm .020$  & $0.106 \pm .005$ \\
$\sigma_{8}$   & & 
  & $0.871 \pm .054$  & $0.857 \pm .059$  &  &  \\
\hline
\hline
+ running &   &  &  &  &  & \\
\hline 
$\Omega_b{\rm h}^2$ & $.0241 \pm .0023$  
 & $.0224 \pm .00162$ & $.0221 \pm .00130$ & $.0221 \pm .00120$ &  $.02289 \pm .00074$  & $.02258 \pm .00017$ \\
$\Omega_{cdm}{\rm h}^2$  & $.122 \pm .0197$ 
& $.121 \pm .0158$  & $.125 \pm .0101$ & $.127 \pm .0087$ & 
$.1113 \pm .0070$  & $.1165 \pm .0014$ \\
$n_s(k_n)$   & $0.968 \pm .076$  &
 $0.906 \pm .0555$  & $0.904 \pm .0456$ & $0.902 \pm .0437$ & $0.976 \pm .0346$ & $0.9595 \pm .0038$ \\
$-dn_s/d\ln k$   & $0.077 \pm .044$ &
$0.0847 \pm .033$ & $0.0736 \pm .031$ & $0.0697 \pm .029$ &
$-0.0463 \pm .025$ & $0.0037 \pm .005$  \\
$h$  & $0.760\pm 0.088$  &
$0.698 \pm 0.075 $ & $0.677 \pm 0.050 $  & $0.669 \pm 0.039 $ & $0.7366 \pm 0.0365$ & $0.7061 \pm 0.0068$  \\
$\tau_C$ & $0.325 \pm .129$  &
$0.248 \pm 0.112$  & $0.213 \pm 0.0926$ & $0.200 \pm 0.0742$ &  $0.0984 \pm .019$ & $0.1083 \pm .0062$ \\
$\sigma_{8}$   & & 
  & $0.935 \pm .066$  & $0.936 \pm .0710$ &  &   \\
\hline
\end{tabular}
\end{center}
\end{sidewaystable}

Fig.~\ref{fig:cbialldnLSS} and Table~\ref{tab:exptparamsMCMC} address
the level at which the WMAP1 and Jun03 CMB data would prefer a running
spectral index $dn_s/d\ln k$ at about the two-sigma level.  The
projected distributions depend upon prior choices, \eg
Fig.~\ref{fig:cbialldnLSS} shows that if one restricts how large
$\tau_C$ is allowed to be, $-dn_s/d\ln k$ would not be as large,
accounting for the differences in the result here with those of the
WMAP1 team's analysis (Spergel \etal 2003). They highlighted how the
non-CMB Lyman-alpha forest data in conjunction with the CMB suggested
a running index. More effort is required to demonstrate that the
forest estimates of power spectra are reliable enough to apply to this
problem.

\begin{figure}
\centerline{\hspace{0.1in}
\epsfxsize=6.0in\epsfbox{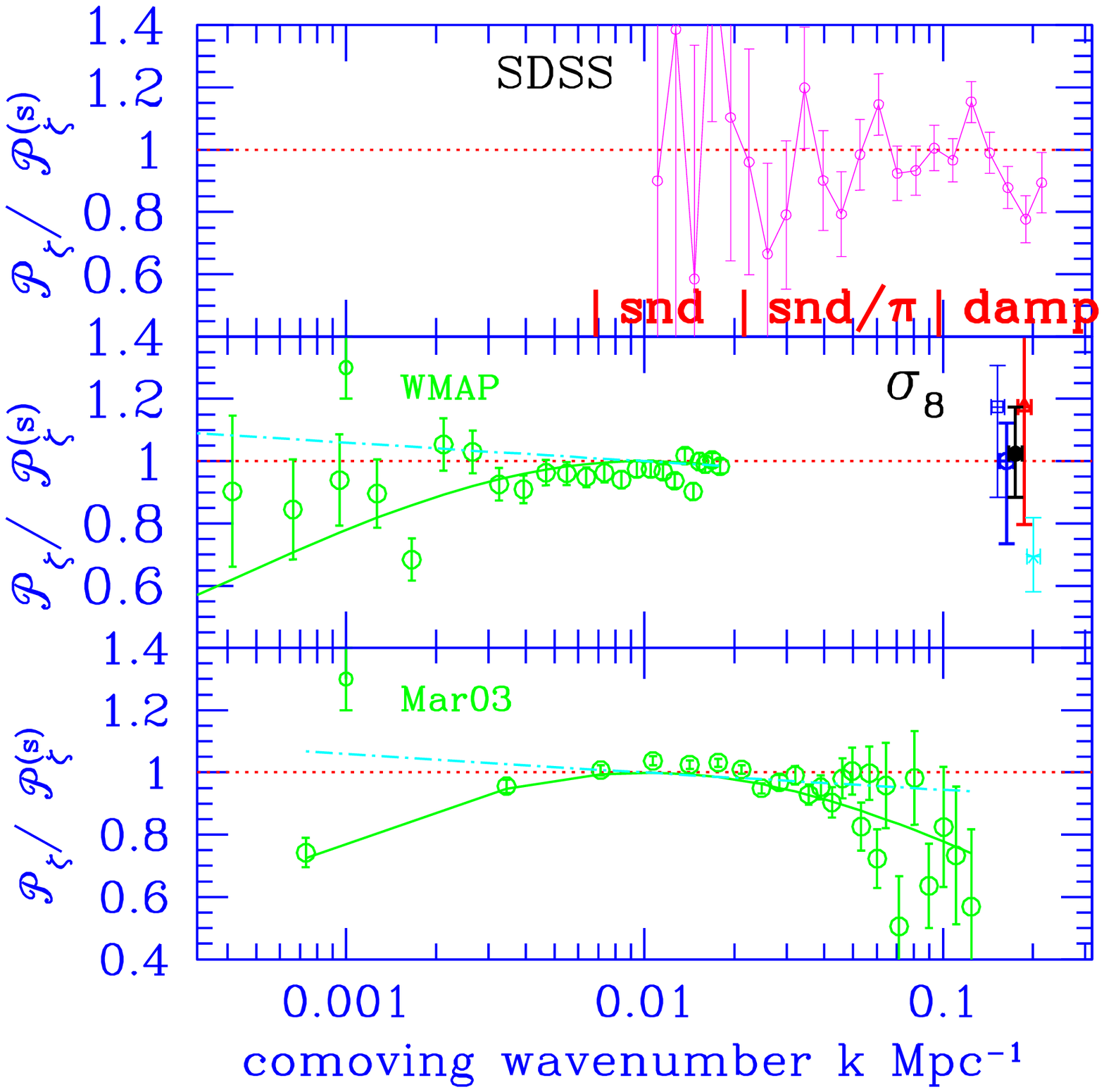} }
\vspace{-0.5in} 
\caption{The bottom panel translates the ${\cal C}_\ell$ spectrum of
Fig.~\ref{fig:CLoptjun03} by mapping it from $\ell$ into
(perpendicular) spatial wavenumber $k_\perp$, where $\ell \sim {\cal
R}_{dec} k_\perp$, and also dividing out the target spectrum. It
indicates the ratio of primordial observed ${\cal P}_\zeta (k) $ to
target ${\cal P}_\zeta^{(s)}(k)$ in $k$-space with a window function
for each band. The window spillover into neighbouring bands is not
large, so it shows where the anomalies lie for the specific best-fit
$n_s$=0.957 $\Lambda$CDM model. The downward bending curve is a shape
that a running index of $-0.09$ would give.  The low and high $\ell$
downward-drive is evident. Breaking up the single low $\ell$ bandpower
into many bandpowers as shown in the middle panel highlights the lower
$\ell$ and $\ell \sim 20$ ``anomalies''. The right side of the middle
panel shows $\sigma_8^2$, a broad-band power that probes cluster
scales which reside near $k^{-1} \sim 4\hmpc$, relative to the target
model. The solid square is derived from the Jun03 data, with $\sigma_8
\approx 0.85$. The heavy open square to its left is the current best
value for weak lensing (Hoekstra and van Waerbeke, 2004, private
communication), which has evolved slightly downward from the Jun03
estimate shown as the leftmost light square. Many X-ray cluster
estimates are lower (rightmost light butterfly). The heavy butterfly
to its immediate left is the $\sigma_8$ estimate from the SZ
interpretation of the CBI, Acbar, BIMA high $\ell$ ``anomaly'' (\eg
Readhead \etal 2004). The top panel shows the recent SDSS data of
Tegmark \etal (2004). Instead of dividing our Jun03 best-fit, which
has some wiggles induced by the baryons, a best-fit wiggle-less
``$\Gamma$-model'' was used to highlight any obvious need, within the
errors: the statistical answer is no (Tegmark \etal 2004).  Galaxy
biasing complicates the use of SDSS and 2dFRS, but there is no
indication of any running index within these LSS datasets.}
\label{fig:pzeta}
\end{figure}

Of course in statistics one should just ignore two sigma indications,
and especially here when an extra cosmic parameter is added which has
strong degeneracies with others in the basic minimal inflation mix.
Fig.~\ref{fig:pzeta} shows the conspiracy driving the indications of
running index from the Jun03 data alone. The two low $\ell$ anomalies
and the slightly lower power at high $\ell$ would prefer to bend the
best-fit uniform acceleration model downward. By itself, WMAP1 does
this, and the addition of other experiments just takes this tendency
and adds to it. However, BCP showed that the pre-WMAP1 Jan03
compilation of the data that included DMR and Archeops also had a
distribution that preferred negative $[dn_s/d\ln k (k_n)]$, though
with less statistical significance than the post-WMAP1 Mar03 set.

Because theorists like to theorize about low significance results in
anticipation they might eventually emerge at high significance, much
renewed discussion and many papers have now been written on whether
the low $\ell$ anomalies or the combination of low and high $\ell$
anomalies indicate new physics. Within the context of inflation
models, this involves arranging for $q$ to change.  If only low $\ell$
is the target, a scale is built associated with a target $k$, \eg
Bridle \etal (2003), Contaldi \etal (2003). Topology is another
mechanism, building in a characteristic horizon-scale size to
discretize $k$-space, with just enough inflation to make the Universe
just so big but no bigger. Trying to solve the high and low $\ell$
anomalies with the same mechanism utilizes the running index, or would
need to build in a mix of scales.

Given this baroqueness, it is useful to explore the sensitivity of the
running index distributions to cuts in the data. For example, Bridle
\etal (2003) found $[dn_s/d\ln k (k_n)]$ of $-0.04 \pm 0.03$ using all
multipoles of WMAP1, the Jun02 versions of CBI and VSA and the Jan03
version of ACBAR, along with 2dF. When they excluded $\ell <5$ from
WMAP1, this dropped to $-0.015 \pm 0.03$. To test this sensitivity
further, we have marginalized over the $\ell = 2,3,4,21,22,33$
multipoles of the WMAP1 data, which have ``anomalous'' bandpowers. We
find $-0.062 \pm 0.043$ compared with $-0.088 \pm 0.041$ with no such
cuts. Just removing $\ell = 22$ gives $-0.082 \pm 0.042$. For this
exercise, we did not constrain the running index by any cosmological
priors and allowed it to vary between $\pm 0.3$, which leads to a
broadened $\tau_C$ distribution, as reflected in
Table~\ref{tab:exptparamsMCMC}.

Of course there should be a reason to justify removing or cleaning
anomalies, \eg that a source of systematic error is found or
foregrounds and other residuals contaminate.  In spite of much debate,
there is no evidence that the low $\ell$'s are low because of
foreground contamination. However, when the quadrupole, octupole,
\etc, are obtained, the influence of the foregrounds should be
reflected in the error bars. Slosar \etal (2004) improved the
determination of errors on WMAP1 at low $\ell$ by marginalizing over
foregrounds rather than using template subtractions. This leads to
more power at low $\ell$ and better error determination.  They improved 
the treatment of likelihood tails at low-$\ell$ over the ``standard
WMAP1 prescription'' of Verde \etal (2003), which we used to get
the results given in our figures and tables. Both effects decrease the
statistical desire for a downturn and Slosar \etal find that the
$\sim$ 2-$\sigma$ effect drops to a $\sim$ 1-$\sigma$ effect.

Our conclusion, as in Bridle \etal (2003), BCP, Readhead \etal (2004)
and Slosar \etal (2004), is that evidence for a running index in the
CMB data is not compelling. To get the large values allowed by the
data would require rather dramatic changes in the acceleration of the
universe over what is actually quite a narrow range on the inflaton
potential surface, manifested in a soft or even a radical breaking
because of changing braking. 

\subsection{Forecasts of Parameter Precision} \label{sec:paramfuture} 

\begin{figure}
\centerline{\hspace{0.1in}
\epsfxsize=6.0in\epsfbox{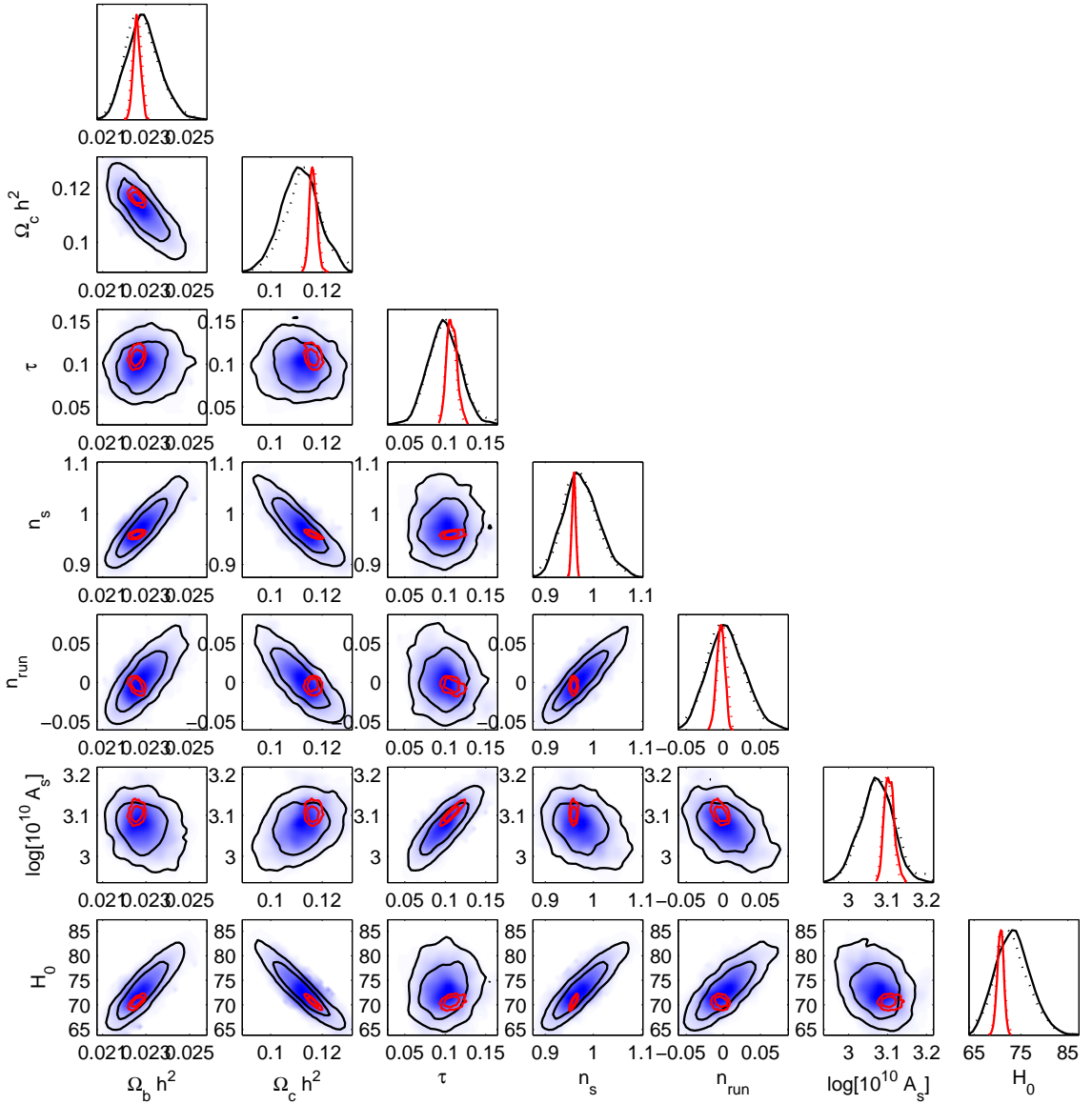} }
\caption{Forecast of one and two sigma contour regions for WMAP4
(black outer contours, light blue shading) and Planck1 (red inner
contours) show how the errors of Fig.~\ref{fig:cbialldnLSS} may
improve in the future. Note some of the variables shown differ
from those in Fig.~\ref{fig:cbialldnLSS} and there are changes in axes
scales.  Estimations were on simulated datasets generated using the
no-running-index best-fit to the Jun03 CMB data. Only the 94 GHz
channels were used for the 4 year simulation of WMAP4 and only the 150
GHz channels for the 1 year simulation of Planck1 in these forecasts.
No LSS priors were imposed.  Table~\ref{tab:exptparamsMCMC} gives
means and errors for these two cases. The precision sharpens when all
channels are brought to bear. Further, both Planck and WMAP are
expected to observe for roughly double these periods, decreasing the
noise component of the bandpower errors, with the sample-variance
(cosmic-variance) component unaffected.  One may therefore interpret
these as conservative estimates, but the forecasts here do not include
all of the extra complications associated with foreground separation.
As Fig,~\ref{fig:fcastrun} shows, anticipated ground-based experiments
beyond the ones used in Fig.~\ref{fig:cbialldnLSS} will also have a
powerful sharpening effect on precision.}
\label{fig:mcmcrun}
\end{figure}

\begin{figure}
\centerline{\hspace{0.1in}
\epsfxsize=6.0in\epsfbox{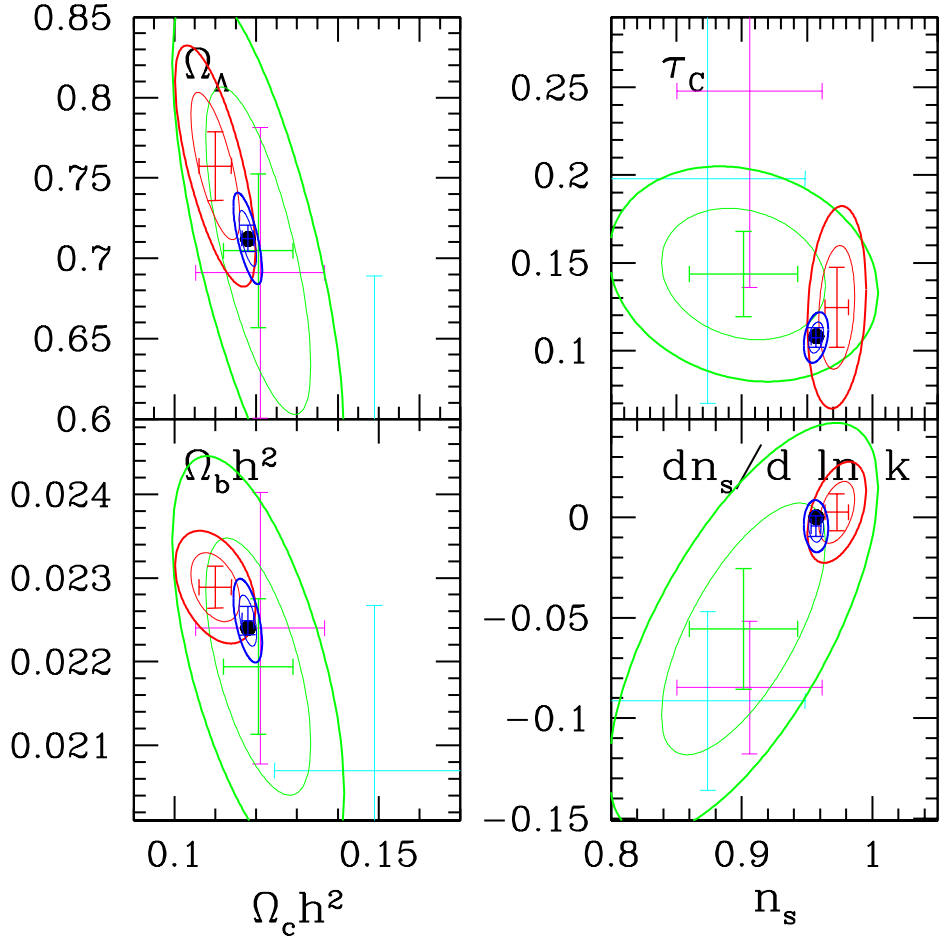} }
\caption{The forecasts of WMAP4 (green), Planck1 (blue) and WMAP4 +
ground-based ACT/SPT-like data (red) are shown compared with the
target value (black dot). Projected one-sigma error bars are shown,
and the one and two sigma ellipses which illustrate the
correlations. The (magenta) cross shows the precision of the Jun03
data when the scalar spectral index is allowed to run, as it is for
the simulations as well. The bigger (cyan) cross shows the state with
the pre-WMAP Jan03 data.  Large dedicated ground-based telescopes
targeting high $\ell$ with huge arrays of bolometers (\eg ACT and SPT)
or of HEMTs (QUIET), when combined with WMAP, should greatly increase
parameter precision in the leadup to Planck. For this simulation,
ACT/SPT experimental parameters were adopted, and the bolometers were
assumed to be polarization sensitive to show their powerful impact on
EE mode detection. The assumed coverage was 2.4\% of the
sky, 1000 square degrees. Increasing this would further improve the
parameter estimates given in Table~\ref{tab:exptparamsforecast2} since
errors on many bands are sample-variance limited.  Planck does so well
because of its all-sky coverage, and will have a large impact 
on constraining multi-parameter deviations from the simple
uniform-acceleration inflation models. }
\label{fig:fcastrun}
\end{figure}

The running index issue will probably be with us for quite a while,
but forecasts are rosy for how well planned CMB experiments can answer
this question: if there is a running index, it will be detected in the
next generation of experiments, and if there is not it will be
strongly constrained. This is illustrated in Fig.~\ref{fig:fcastrun},
Table~\ref{tab:exptparamsforecast2} as well as in
Fig.~\ref{fig:mcmcrun} and Table~\ref{tab:exptparamsMCMC}.

The precision for ground + WMAP4 would improve with larger sky
coverage.  For the numbers in the table, it was assumed that the
primary spectrum beyond 2000 would be contaminated by secondary
signals, but component separation should mitigate this, and the
parameters are not sensitive to lesser cutting. Note how the
larger baseline in $\ell$ significantly decreases the degeneracy and
also the drift in the value of the running index one would estimate
from the data relative to the target value.

For the forecasts of Table~\ref{tab:exptparamsforecast2}, power
spectra and their errors in Figs.~\ref{fig:pol4_wmap4},
\ref{fig:pol4_planck1}, \ref{fig:pol4_wmap4_SPT1000} were calculated
using ``faster-like'' algorithms of the sort we have appled to
Boomerang. The idealizations make it ``superfast''.  The cosmic
parameter errors were estimated using Fisher or curvature matrices
about maximum likelihood values, rather than using Monte Carlo Markov
Chains on the simulated data as in
Table~\ref{tab:exptparamsMCMC}. This means the parameters are treated
as completely internal rather than completely external, as in MCMC.
(See BCP for a discussion of the difference; the fixed grid approach
was a hybrid, with amplitudes and experimental variables treated as
internal, the rest external.)  For both forecasting methods, the same
Jun03 best-fit model with zero running index was used and the
experimental parameters were essentially the same. The maximum
likelihood drifts from the listed input cosmic parameter values of the
target model depend upon the specific realization. Note that the
errors of the Fisher-based forecast given here are quite similar to
the MCMC values given in the last two columns of
Table~\ref{tab:exptparamsMCMC}. 

Thus the superfast forecasts have been nicely validated by comparison
with the MCMC results. They also give results compatible with what was
actually obtained with Boomerang and WMAP when the real experimental
specs were used. The forecasts can be made more sophisticated with
some attempts at addition of foregrounds and residual signals and
subsequent removal by parameterized ``power spectrum
cleaning''. Tegmark \etal (2000) explored many aspects of the impact
foreground contamination could have on forecasted cosmological
parameter errors. Increasingly sophisticated forecasts can still only
be considered as partial steps towards the full mocks of a given
experiment that one actually needs.

Forecasting has a long history in the CMB, as a necessary ingredient
for experimental proposals, and for showing feasibility of measuring
new theoretical effects.  Many realizations, experimental
configurations and theoretical assumptions can be checked very
quickly. This leads to an overwhelming amount of information on the
stages we expect to see between the data now and the Planck results,
because there are so many polarization-sensitive ground-based experiments
in various configurations either funded or proposed. Further there are
many additional theoretical parameters to consider to further add to the
information glut, yet turning them all on at the same time obscures
what will happen in practice,

In this paper, we have chosen to highlight only three forecasting
cases, WMAP4, Planck and a fiducial high resolution ground-based
experiment of modest sky coverage compared with what is possible. This
reflects our experience that erring on the conservative side may
reflect the real issue, which is how complex the analysis of the
actual polarization data will turn out to be, and how much it will
limit the precision we can obtain from the ground; and indeed from
space. For polarization-targeting ground experiments, we can look
forward to a wonderful set of developments. Listing some of the cases
that have been considered gives an idea of the range.

Apart from WMAP4 and Planck1 with one channel, we have considered the
following: WMAP2, WMAP4, WMAP8, using all 5 channels as well as the W
channel with its $13^\prime$ resolution adopted here, usually assuming
0.9 for $f_{sky}$ -- detector noise and beam sizes courtesy of Lyman
Page; Planck1 and Planck2, using either the 143 GHz channel with beam
$7^\prime$ alone, or with the 220 GHz, $5^\prime$ and 100 GHz,
$9^\prime$ polarization-sensitive bolometers (PSBs), or all together with
the lower frequency HEMTs -- from Planck ``blue book'' numbers,
augmented by the most recent PSB numbers from Andrew Lange. With
everything included forecasted errors do improve somewhat, but are often
largely sample variance limited. 

We have considered forecasts for the Boomerang 2003 flight (the
polarization analysis of the real data is heavily underway), and for
Acbar (which is continuing to observe) -- specs for current and
subsequent observing seasons from John Ruhl.

We have also forecasted for the South-Pole-based BICEP, using PSBs at
143 GHz, $40^\prime$ and 100 GHz, $60^\prime$ observing 1000 sq deg in
260 days -- numbers from Lange and Eric Hivon, with similar
capabilities suggested for an experiment at another Antarctic site,
Dome C, courtesy of Paolo de Bernardis. For QUaD (Quest mounted on
DASI at the South Pole), we used PSBs at 143 GHz, $4.0^\prime$ and 100
GHz, $6.3^\prime$ observing 200 sq deg in 260 days -- specs from Lange
and Hivon, Similar numbers are given for a separate Cardiff-based
proposal. Prospects for BICEP and QUaD are very good for polarization 
and both are expected to be observing in 2005. See Bowden
et al. (2004) for a full discussion of optimizing ground-based CMB
polarization experiments, in terms of the tradeoff of sky coverage and
sensitivity per pixel.

The proposed QUIET experiment from Princeton, Chicago, JPL/Caltech,
using new HEMT-MMIC array technology under development at JPL, would
be mounted on the CBI platform in Chile, with large beams, 44 and 90
GHz, $42^\prime$, over 2000 and 8000 sq deg, and small beams, 44 and
90 GHz, $4^\prime$, over 2000 and 8000 sq deg. Forecasts look very
good for polarization and the first phase could begin in late 2005. The
South Pole Telescope numbers used only 220 GHz, $1.3^\prime$, the SZ
null channels -- were from John Ruhl. We assumed a noise for
polarization $\sqrt{2}$ times that in total anisotropy per pixel.  For
the Chile-based ACT, the beam is slightly larger than for the SPT. A
deep mode of 100 sq deg was considered in addition to the 1000 sq deg
we chose to highlight here -- sensitivities from Lyman Page. In
contrast to Fig.~\ref{fig:pol4_wmap4_SPT1000}, this showed the power
in the lensing-induced B-mode could in principle be detected.  A
fiducial CMBPol and an essentially cosmic variance limited all-sky
survey at SPT/ACT resolution with very tiny noise have also been
considered, Needless to say, for the latter the target parameters are
recovered at the best-you-can-do level.

We now turn back to the results shown in
Table~\ref{tab:exptparamsforecast2}.  The last set of rows show how
the error bars open up when searching for more radical braking than
the running index model gives. Consider the case when ${\cal P}_\zeta
(k)$ has a structure of unknown shape, as in radical broken scale
invariance (BSI). For two given cosmological parameter sets, a ${\cal
P}_\zeta (k)$ could be fashioned to morph one ${\cal C}_\ell^{(TT)}$
into another. (See \eg Souradeep \etal (1998), Wang \etal (1999) for a
discussion of the role this degeneracy plays in parameter
degradation.)  Polarization information breaks this severe degeneracy
because the peaks and dips of ${\cal C}_\ell^{(EE)}$ and ${\cal
C}_\ell^{(TT)}$ are in different locales. Parameterization is in terms
of power amplitudes in a number of $k$-bands of proscribed shape. 24
bands were chosen for the Table~\ref{tab:exptparamsforecast2} case.
Apart from the conventional banding in $\Delta P_\zeta$, we have
expored the impact of band-colours (bands in $\Delta n_s$), continuous
wavelets, among others.  The colour-banding makes more of a
difference. However, although the errors determined are somewhat
sensitive to the primordial spectrum band-type, band-placing and
band-number, polarization does indeed nicely mitigate the effect of
BSI-induced degeneracy for these planned experiments. Non-CMB
information from LSS also helps to break degeneracies between cosmic
parameters and ${\cal P}_\zeta (k)$-structure. Having $\tau_C$ from
the low $\ell$ is important for breaking parameter degeneracies.

As more parameters characterizing the inflation model are added, the
precision continues to diminish unless near-degeneracies can be
broken. For example, with a target value for ${\cal P}_{GW}/{\cal
P}_\zeta$ of 0.17, the Planck1 realization of
Fig.~\ref{fig:pol4_planck1} shows a detection of the tensor B-mode is
possible, which could lead to a good estimate of this
amplitude. Getting the B-mode is very important for this. The specific
example, determining 8 other cosmological parameters as well, gave a
$0.135\pm 0.028$ detection.  With somewhat more optimistic noise
forecasts, but allowing for the incomplete sky coverage mixing of E
and B modes, Lewis (2003) finds that Planck should be able to detect
primordial tensor modes at 95\% confidence with greater than 95\%
probability if ${\cal P}_{GW}/{\cal P}_\zeta \gta 0.03$.

We can conclude from exercises such as these on the experiments coming
that parameters characterizing GW signals, mildly broken scale invariance
associated with a running index, subdominant isocurvature
components, and even radically broken scale invariance can be
determined within the CMB data.

\begin{table}
\caption{Forecasted cosmological parameter values and their 1-sigma
errors for the cases of Fig.~\ref{fig:fcastrun} contrast what may be
achievable for WMAP4, Planck1, and for future ground-based data
(labeled ACT/SPT). The last column combines a polarization-sensitive
bolometer-array ACT/SPT-like telescope (experimental choices as in
Fig,~\ref{fig:fcastrun}) with WMAP4.  Increasing the assumed coverage
beyond the adopted 2.4\% of the sky would increase precision in these
idealized forecasts.  For WMAP4 and Planck1, the quadrupole was
included and the useful sky coverage was chosen to be unity. When the
quadrupole is excluded, the sky coverage is dropped to 90\%,
$\omega_{m\nu}$ and ${\cal P}_{GW}/{\cal P}_{\zeta}$ are added to the
parameter mix, and a tensor component and weak lensing effects on the
power spectra are included in the target model, the errors grow
modestly, usually within 20\% of the sigmas listed. Of the 9
parameters, WMAP4 would determine 3 orthogonal combinations to $\pm
0.01$, 7 to $\pm 0.1$; with an SPT/ACT-like experiment, these rise to
4 and 8; and for Planck1, to 6 and 8. The lower rows illustrate the
impact of including many more parameters to characterize $n_s (k)$
than the two in the running index model.  Here 24 parameters defining
the amplitudes of ${\cal P}_\zeta (k)$ in 24 bands are added to the
standard mix. The impact is not as severe as one might have expected
since the polarization information breaks this severe degeneracy
because the peaks and dips of ${\cal C}_\ell^{(EE)}$ and ${\cal
C}_\ell^{(TE)}$ are in different locales than ${\cal
C}_\ell^{(TT)}$. In Souradeep \etal (1998), it was shown that
parameter determination was significantly degraded if there was no
polarization information when similar numbers of ${\cal P}_\zeta
(k)$ bands were added. }
\label{tab:exptparamsforecast2}
\begin{center}
\begin{tabular}{|ll|l|l|l|}
\hline
 & input    & WMAP4 & Planck1 & WMAP4+ACT/SPT-like \\
\hline
flat    &  &  & &  \\
\hline 
$\Omega_b{\rm h}^2$ & $.02240$ & $.02311 \pm .00051$  &$.02243 \pm .00015$ &$.02291 \pm .00025$ 
 \\
$\Omega_{cdm}{\rm h}^2$ & $.1180$   & $.1077 \pm .0049$  & $.1181 \pm .0014$ & $.1105 \pm .0035$  
 \\
$n_s$ & $.9570$   &  $0.9738 \pm .0137$ &  $0.9567 \pm .0037$ & $0.9716 \pm .0080$   \\
$\tau_C$ & $.1080$ & $0.1274 \pm .023$   & $0.106 \pm .005$  & $0.1270 \pm .021$   \\
\hline
\hline
+ running   &  &  & &   \\
\hline 
$\Omega_b{\rm h}^2$  & $.02240$ &  $.02194 \pm .00081$  & $.02236 \pm .00017$& $.02289 \pm .00025$ \\
$\Omega_{cdm}{\rm h}^2$   & $.1180$  & $.1205 \pm
.0085$  & $.1166 \pm .0015$ & $.1100 \pm .0040$ \\
$n_s(k_n)$  & $.9570$    & $0.9014 \pm .0415$ & $0.9569 \pm .0038$ & $0.9727 \pm .0090$ \\
$-dn_s/d\ln k (k_n)$ & $0$    & $-0.0555 \pm .030$ & $ -0.0044 \pm .0052$ & $ 0.0025 \pm .0092$  \\
$\tau_C$ & $.1080$ &  $0.1436 \pm .024$  & $0.1074 \pm .0056$ & $0.1246 \pm .0229$  \\
\hline
\hline
+ BSI    &  &  & &   \\
\hline 
$\Omega_b{\rm h}^2$  & $.02240$ &  $.02974 \pm .00196$  & $.02223 \pm .00022$& $.02266 \pm .00042$ \\
$\Omega_{cdm}{\rm h}^2$   & $.1180$  & $.1059 \pm
.0168$  & $.1192\pm .0020$ & $.1067 \pm .0057$ \\
$n_s $  & $.9570$    & $1.183 \pm .074$ & $0.9709 \pm .0146$ & $0.9688 \pm .0388$ \\
$\tau_C$ & $.1080$  &  $0.2034 \pm .030$ & $0.1123 \pm .0051$ & $0.1602 \pm .0266$  \\
\hline
\end{tabular}
\end{center}
\end{table}


\begin{thebibliography}{}

\bibitem{bon96} Bond, J.R. 1996, {\it Theory and Observations of the
Cosmic Background Radiation}, in ``Cosmology and Large Scale
Structure'', Les Houches Session LX, August 1993, ed.  R. Schaeffer,
Elsevier Science Press.

\bibitem{yukawa93} J.R. Bond, 1994, { Testing inflation with the cosmic
background radiation}, in: M. Sasaki, ed., { Relativistic Cosmology},
Proc. 8th Nishinomiya-Yukawa Memorial Symposium (Academic Press)
astro-ph/9406075.


\bibitem{BCP03} Bond, J.R., Contaldi, C.R. \& Pogosyan, D. 2003,
%``Cosmic microwave background snapshots: pre-WMAP and post-WMAP,''
Phil.\ Trans.\ Roy.\ Soc.\ Lond.\ A 361, 2435, astro-ph/0310735,
BCP 

\bibitem{Bond02} Bond, J.R., et al.~2004, ApJ, in  press, astro-ph/0205386

\bibitem{BC01} Bond, J.R.\ \& Crittenden, R.G. 2001, in Proc.\ NATO ASI,
Structure Formation in the Universe, eds.\ R.G.\ Crittenden \& 
N.G.\ Turok (Kluwer), astro-ph/0108204

\bibitem{bowden04} Bowden, M. et al. 2004, MNRAS, Volume 349, Issue 1, pp. 321-335, astro-ph/0309610
 
\bibitem{bridle03} Bridle, S.L.,  Lewis, A.M., Weller, J. \& Efstathiou, G. 2003, MNRAS
342, L72, astro-ph/0302306

\bibitem{contaldi03} Contaldi, C.R. et al. 2003, JCAP 0307, astro-ph/0303636 

\bibitem{vsa04} Dickinson, C. et al. 2004, submitted to MNRAS, astro-ph/0402498

\bibitem{Goldstein03} Goldstein, J.H.~et  al.\ 2003, \apj 599, 773, astro-ph/0212517 

\bibitem{habib02} Habib, S.,  Heitmann, K., Jungman, G. \& Molina-Paris, C. 2002, Phys. Rev. Lett. 89, 281301
     
\bibitem{hivon02} Hivon, E. \& Kamionkowksi, M. 2002 {Science} 298, 1349. astro-ph/0211553 

\bibitem{hu00} Hu, W. 2000, Phys. Rev. D62, 043007

\bibitem{kinney02} Kinney, W.H. 2002, Phys. Rev. D66, 083508

\bibitem{kl87} Kofman, L.A. \& Linde, A.D. 1987, Nucl. Phys. {B282}, 555; Kofman, L.A.  \& Pogosyan, D.Yu. 1988 {Phys. Lett.}
{214 B}, 508

\bibitem{leachliddle03} Leach, S.M. \& Liddle, A.R. 2003, Phys. Rev. D68, 103503, astro-ph/0306305

\bibitem{lewis03}  Lewis, A.M. 2003, Phys. Rev. D68, 083509,   astro-ph/0305545 
 
\bibitem{l2kcba97} Lidsey, J.E.  et al. 1997, Rev. Mod. Phys. {69} 373

\bibitem{martin03}  Martin, J. \& Schwarz, D.J. 2003, Phys. Rev. D67, 083512, astro-ph/0210090

\bibitem{pagepkdip} Page, L.~et al. 2003,  ApJSupp 148, 233

\bibitem{peiris} Peiris, H.V.~et al. 2003,  ApJSupp 148, 213

\bibitem{cbi7} Readhead, A.C.S.et al. 2004, ApJ, in press, astro-ph/0402359 

\bibitem{sbb}
Salopek, D.S., Bond, J.R. \& Bardeen, J.M. 1989, \prd 40, 1753

\bibitem{sb12} Salopek, D.S. \& Bond, J.R. 1990, \prd {42},
3936; 1991 \prd {43}, 1005

\bibitem{schwarz01} Schwarz, D.J., Terrero-Escalante, C.A., Garcia,
  A.A. 2001, Phys. Lett. B517, 243

\bibitem{spergel03} Spergel, D.~N.~et al. 2003, ApJSupp 148, 175 

\bibitem{slosar} Slosar, A., Seljak, U. \& Makarov, A. 2004, astro-ph/0403073

\bibitem{ambleside98} Souradeep, T., Bond, J.R., Knox, L., Efstathiou,
G. \& Turner, M.S. 1998, {\it Prospects for measuring Inflation
parameters with the CMB}, in ``Proceedings of COSMO-97'', Ambleside,
UK, Sept. 1997, ed. L. Roszkowski (World
Scientific), astro-ph/9802262 

\bibitem{tegmark00} Tegmark, M., Eisenstein, D.J., Hu, W., de Olivera-Costa, A. 2000, ApJ
530, 133

\bibitem{sdss} Tegmark, M. et al. 2004, ApJ, in press, astro-ph/0310725 

\bibitem{verde03} Verde, L.~et al. 2003, ApJSupp 148, 195

\bibitem{wang97} Wang, L., Mukhanov, V.F. \& Steinhardt, P.J.  1997, Phys. Lett. B414,  18

\bibitem{spergel99} Wang, Y., Spergel, D.N. \& Strauss,  M.A.  1999, ApJ 510, 20

\end{thebibliography}
\end{document}